\documentclass[twocolumn]{aastex62}

\newcommand{\hatcur}{HATS-70}
\newcommand{\hatcurb}{HATS-70b}

\newcommand{\genevastarmass}{\ensuremath{1.78\pm0.12}}                 
\newcommand{\genevastarradius}{\ensuremath{1.881_{-0.066}^{+0.059}}}                 

\newcommand{\genevastarteff}{\ensuremath{7930_{-820}^{+630}}}              
\newcommand{\genevastarlogg}{\ensuremath{4.167_{-0.036}^{+0.044}}}              
\newcommand{\genevastarfeh}{\ensuremath{0.041_{-0.107}^{+0.095}}}              
\newcommand{\genevastarlum}{\ensuremath{12.0_{-3.4}^{+5.5}}}              
\newcommand{\genevastarage}{\ensuremath{0.81_{-0.33}^{+0.50}}}              

\newcommand{\genevastarvsini}{\ensuremath{40.61_{-0.35}^{+0.32}}}             
\newcommand{\genevastarvmac}{\ensuremath{6.16_{-0.67}^{+0.70}}}             
\newcommand{\genevaebmv}{\ensuremath{< 0.069 \, (1\sigma)}}              
\newcommand{\genevairot}{\ensuremath{> 74.9\, (1\sigma)}}              
\newcommand{\genevastaroblate}{\ensuremath{0.98561_{-0.00072}^{+0.00066}}}              

\newcommand{\genevatzero}{\ensuremath{2456911.87486_{-0.00088}^{+0.00080}}}              
\newcommand{\genevaperiodshort}{\ensuremath{1.89}}              
\newcommand{\genevaperiod}{\ensuremath{1.8882378\pm0.0000015}}              
\newcommand{\genevarratio}{\ensuremath{0.0740\pm0.0028}}              
\newcommand{\genevainc}{\ensuremath{86.7_{-1.9}^{+1.6}}}              
\newcommand{\genevalambda}{\ensuremath{8.9_{-4.5}^{+5.6}}}              

\newcommand{\genevartecosw}{\ensuremath{0.01_{-0.14}^{+0.12}}}              
\newcommand{\genevartesinw}{\ensuremath{0.34_{-0.29}^{+0.16}}}              
\newcommand{\genevaecc}{\ensuremath{<0.18 \, (1\sigma)}}              
\newcommand{\genevajitterA}{\ensuremath{440_{-300}^{+530}}}              
\newcommand{\genevajitterB}{\ensuremath{620_{-140}^{+170}}}              
\newcommand{\genevagammaA}{\ensuremath{36.22_{-0.42}^{+0.41}}}              
\newcommand{\genevagammaB}{\ensuremath{35.75_{-0.20}^{0.19}}}              

\newcommand{\genevaKrv}{\ensuremath{1460_{-190}^{+220}}}              

\newcommand{\genevaa}{\ensuremath{0.03632_{-0.00087}^{+0.00074}}}              
\newcommand{\genevaars}{\ensuremath{4.17_{-0.13}^{+0.16}}}              
\newcommand{\genevab}{\ensuremath{0.24_{-0.11}^{+0.13}}}              
\newcommand{\genevatdur}{\ensuremath{0.1510_{-0.0048}^{+0.0049}}}              

\newcommand{\genevaplanetmass}{\ensuremath{12.9_{-1.6}^{+1.8}}}                 
\newcommand{\genevaplanetradius}{\ensuremath{1.384_{-0.074}^{+0.079}}}                 
\newcommand{\genevaplanetrho}{\ensuremath{6.64_{-1.3}^{+1.6}}}         
\newcommand{\genevaplanetgrav}{\ensuremath{4.250_{-0.079}^{+0.075}}}         
\newcommand{\genevaplanetteq}{\ensuremath{2730_{-160}^{+140}}}         
\newcommand{\genevaplanetFincident}{\ensuremath{1.24\pm0.023 \times 10^{10}}}         
\newcommand{\genevaplanetSaf}{\ensuremath{5.842_{-0.080}^{+0.087} \times 10^{12}}}         

\newcommand{\dartmouthstarmass}{\ensuremath{1.683_{-0.032}^{+0.018}}}                 
\newcommand{\dartmouthstarradius}{\ensuremath{1.886_{-0.052}^{+0.045}}}                 

\newcommand{\dartmouthstarteff}{\ensuremath{7420_{-150}^{+160}}}              
\newcommand{\dartmouthstarlogg}{\ensuremath{4.115_{-0.040}^{+0.050}}}              
\newcommand{\dartmouthstarfeh}{\ensuremath{0.006_{-0.099}^{+0.082}}}              
\newcommand{\dartmouthstarlum}{\ensuremath{9.1_{-0.7}^{+1.1}}}              
\newcommand{\dartmouthstarage}{\ensuremath{1.16_{-0.21}^{+0.28}}}              

\newcommand{\dartmouthstarvsini}{\ensuremath{40.63_{-0.33}^{+0.35}}}             
\newcommand{\dartmouthstarvmac}{\ensuremath{6.02_{-0.55}^{+0.57}}}             
\newcommand{\dartmouthebmv}{\ensuremath{< 0.033 \, (1\sigma)}}              
\newcommand{\dartmouthirot}{...}              
\newcommand{\dartmouthstaroblate}{...}              

\newcommand{\dartmouthtzero}{\ensuremath{2456911.87499_{-0.00087}^{+0.00083}}}              
\newcommand{\dartmouthperiod}{\ensuremath{1.8882375\pm0.0000016}}              
\newcommand{\dartmouthrratio}{\ensuremath{0.0724_{-0.0027}^{+0.0026}}}              
\newcommand{\dartmouthinc}{\ensuremath{86.5_{-1.9}^{+1.8}}}              
\newcommand{\dartmouthlambda}{\ensuremath{8.5_{-3.9}^{+6.6}}}              

\newcommand{\dartmouthrtecosw}{\ensuremath{0.09_{-0.26}^{+0.25}}}              
\newcommand{\dartmouthrtesinw}{\ensuremath{0.00_{-0.14}^{+0.12}}}              
\newcommand{\dartmouthecc}{\ensuremath{<0.075 \, (1\sigma)}}              
\newcommand{\dartmouthjitterA}{\ensuremath{396_{-280}^{+480}}}              
\newcommand{\dartmouthjitterB}{\ensuremath{670_{-150}^{+200}}}              
\newcommand{\dartmouthgammaA}{\ensuremath{36.21_{-0.39}^{+0.38}}}              
\newcommand{\dartmouthgammaB}{\ensuremath{35.72_{-0.25}^{0.23}}}              

\newcommand{\dartmouthKrv}{\ensuremath{1461_{-210}^{+240}}}              

\newcommand{\dartmoutha}{\ensuremath{0.03555_{-0.00022}^{+0.00013}}}              
\newcommand{\dartmouthars}{\ensuremath{4.052_{-0.091}^{+0.112}}}              
\newcommand{\dartmouthb}{\ensuremath{0.25_{-0.12}^{+0.13}}}              
\newcommand{\dartmouthtdur}{\ensuremath{0.1549\pm0.0041}}              

\newcommand{\dartmouthplanetmass}{\ensuremath{12.5_{-1.8}^{+2.0}}}                 
\newcommand{\dartmouthplanetradius}{\ensuremath{1.371_{-0.067}^{+0.062}}}                 
\newcommand{\dartmouthplanetrho}{\ensuremath{6.6_{-1.2}^{+1.5}}}         
\newcommand{\dartmouthplanetgrav}{\ensuremath{4.248_{-0.076}^{+0.075}}}         
\newcommand{\dartmouthplanetteq}{\ensuremath{2590\pm20}}         
\newcommand{\dartmouthplanetFincident}{\ensuremath{1.014_{-0.050}^{+0.044} \times 10^{10}}}         
\newcommand{\dartmouthplanetSaf}{\ensuremath{5.94_{-0.089}^{+0.097} \times 10^{12}}}         

\newcommand{\ctbd}[1]{}

\newcommand{\lc}{light curve}

\newcommand{\Lc}{Light curve}

\newcommand{\kms}{\ensuremath{\rm km\,s^{-1}}}
\newcommand{\ms}{\ensuremath{\rm m\,s^{-1}}}

\newcommand{\gcmc}{\ensuremath{\rm g\,cm^{-3}}}
\newcommand{\ergscmsq}{\ensuremath{\rm erg\,s^{-1}\,cm^{-2}}}

\newcommand{\teff}{\ensuremath{T_{\rm eff}}}
\newcommand{\logg}{\ensuremath{\log{g}}}
\newcommand{\vsini}{\ensuremath{v \sin{I_\star}}}
\newcommand{\feh}{\ensuremath{\rm [Fe/H]}}

\newcommand{\rsun}{\ensuremath{R_\sun}}
\newcommand{\msun}{\ensuremath{M_\sun}}
\newcommand{\lsun}{\ensuremath{L_\sun}}

\newcommand{\rstar}{\ensuremath{R_\star}}
\newcommand{\mstar}{\ensuremath{M_\star}}
\newcommand{\lstar}{\ensuremath{L_\star}}

\newcommand{\teffstar}{\ensuremath{T_{\rm eff\star}}}

\newcommand{\loggstar}{\ensuremath{\log{g_{\star}}}}

\newcommand{\rpl}{\ensuremath{R_{p}}}
\newcommand{\mpl}{\ensuremath{M_{p}}}

\newcommand{\rhopl}{\ensuremath{\rho_{p}}}

\newcommand{\arstar}{\ensuremath{a/\rstar}}

\newcommand{\rjup}{\ensuremath{R_{\rm J}}}
\newcommand{\mjup}{\ensuremath{M_{\rm J}}}

\newcommand{\rjuplong}{\ensuremath{R_{\rm Jup}}}
\newcommand{\mjuplong}{\ensuremath{M_{\rm Jup}}}

\graphicspath{{./}{figures/}}

\accepted{\today}
\submitjournal{AJ}

\shorttitle{\hatcurb{}}
\shortauthors{Zhou et al.}

\begin{document}

\title{\hatcur\lowercase{b}: A 13\mjup{} brown dwarf transiting an A star 
\footnote{\scriptsize{The HATSouth network is operated by a collaboration consisting of
Princeton University (PU), the Max Planck Institute f\"ur Astronomie
(MPIA), the Australian National University (ANU), and the Pontificia
Universidad Cat\'olica de Chile (PUC).  The station at Las Campanas
Observatory of the Carnegie Institute is operated by PU in
conjunction with PUC, the station at the High Energy Spectroscopic
Survey (H.E.S.S.) site is operated in conjunction with MPIA, and the
station at Siding Spring Observatory (SSO) is operated jointly with
ANU.
This  paper  includes  data  gathered  with  the  MPG  2.2 m  and
ESO 3.6 m telescopes at the ESO Observatory in La Silla.
This paper includes data gathered with the 6.5~meter Magellan Telescopes located at Las Campanas Observatory, Chile.}
}}

\correspondingauthor{George~Zhou}
\email{george.zhou@cfa.harvard.edu}

\author{G.~Zhou}
\affiliation{Harvard-Smithsonian Center for Astrophysics, 60 Garden St., Cambridge, MA 02138, USA.}
\affiliation{Hubble Fellow}

\author{G.\'A.~Bakos}
\affiliation{Department of Astrophysical Sciences, Princeton University, NJ 08544, USA.}
\affiliation{Packard Fellow}
\affiliation{MTA Distinguished Guest Fellow, Konkoly Observatory, Hungary}

\author{D.~Bayliss}
\affiliation{Dept. of Physics, University of Warwick, Gibbet Hill Road, Coventry CV4 7AL, UK}

\author{J.~Bento}
\affiliation{Research School of Astronomy and Astrophysics, Australian National University, Canberra, ACT 2611, Australia.}

\author{W.~Bhatti}
\affiliation{Department of Astrophysical Sciences, Princeton University, NJ 08544, USA.}

\author{R.~Brahm}
\affiliation{Millennium Institute of Astrophysics, Santiago, Chile}
\affiliation{Instituto de Astrof\'isica, Facultad de F\'isica, Pontificia Universidad Cat\'olica de Chile, Av. Vicu\~na Mackenna 4860, 7820436 Macul, Santiago, Chile.}

\author{Z.~Csubry}
\affiliation{Department of Astrophysical Sciences, Princeton University, NJ 08544, USA.}

\author{N.~Espinoza}
\affiliation{Max-Planck-Institut f\"ur Astronomie, K\"onigstuhl 17, 69117 Heidelberg, Germany.}

\author{J.D.~Hartman}
\affiliation{Department of Astrophysical Sciences, Princeton University, NJ 08544, USA.}

\author{T.~Henning}
\affiliation{Max-Planck-Institut f\"ur Astronomie, K\"onigstuhl 17, 69117 Heidelberg, Germany.}

\author{A.~Jord\'an}
\affiliation{Millennium Institute of Astrophysics, Santiago, Chile}
\affiliation{Instituto de Astrof\'isica, Facultad de F\'isica, Pontificia Universidad Cat\'olica de Chile, Av. Vicu\~na Mackenna 4860, 7820436 Macul, Santiago, Chile.}

\author{L.~Mancini}
\affiliation{Department of Physics, University of Rome Tor Vergata, Via della Ricerca Scientifica 1, I-00133 -- Roma, Italy}
\affiliation{Max-Planck-Institut f\"ur Astronomie, K\"onigstuhl 17, 69117 Heidelberg, Germany.}
\affiliation{INAF -- Astrophysical Observatory of Turin, Via Osservatorio 20, I-10025 -- Pino Torinese, Italy }

\author{K.~Penev}
\affiliation{Physics Department, University of Texas at Dallas, 800 W Campbell Rd. MS WT15, Richardson, TX 75080, USA}

\author{M.~Rabus}
\affiliation{Instituto de Astrof\'isica, Facultad de F\'isica, Pontificia Universidad Cat\'olica de Chile, Av. Vicu\~na Mackenna 4860, 7820436 Macul, Santiago, Chile.}
\affiliation{Max-Planck-Institut f\"ur Astronomie, K\"onigstuhl 17, 69117 Heidelberg, Germany.}

\author{P.~Sarkis}
\affiliation{Max-Planck-Institut f\"ur Astronomie, K\"onigstuhl 17, 69117 Heidelberg, Germany.}

\author{V.~Suc}
\affiliation{Instituto de Astrof\'isica, Facultad de F\'isica, Pontificia Universidad Cat\'olica de Chile, Av. Vicu\~na Mackenna 4860, 7820436 Macul, Santiago, Chile.}

\author{M.~de Val-Borro}
\affiliation{Astrochemistry Laboratory, Goddard Space Flight Center, NASA, 8800 Greenbelt Rd, Greenbelt, MD 20771, USA}

\author{J.E.~Rodriguez}
\affiliation{Harvard-Smithsonian Center for Astrophysics, 60 Garden St., Cambridge, MA 02138, USA.}

\author{D.~Osip}
\affiliation{Las Campanas Observatory, Carnegie Institution of Washington, Colina el Pino, Casilla 601 La Serena, Chile}

\author{L.~Kedziora-Chudczer}
\affiliation{School of Physics, University of New South Wales, Sydney, NSW 2052, Australia}
\affiliation{Australian Centre for Astrobiology, University of New South Wales, Sydney, NSW 2052, Australia}

\author{J.~Bailey}
\affiliation{School of Physics, University of New South Wales, Sydney, NSW 2052, Australia}
\affiliation{Australian Centre for Astrobiology, University of New South Wales, Sydney, NSW 2052, Australia}

\author{C.G.~Tinney}
\affiliation{School of Physics, University of New South Wales, Sydney, NSW 2052, Australia}
\affiliation{Australian Centre for Astrobiology, University of New South Wales, Sydney, NSW 2052, Australia}

\author{S.~Durkan}
\affiliation{Astrophysics Research Centre, Queens University, Belfast, Belfast, Northern Ireland, UK}

\author{J. L\'az\'ar}
\affiliation{Hungarian Astronomical Association, 1451 Budapest, Hungary}

\author{I. Papp}
\affiliation{Hungarian Astronomical Association, 1451 Budapest, Hungary}

\author{P. S\'ari}
\affiliation{Hungarian Astronomical Association, 1451 Budapest, Hungary}




\begin{abstract}
We report the discovery of \hatcurb{}, a transiting brown dwarf at the deuterium burning limit. \hatcurb{} has a mass of $M_p = \genevaplanetmass\,\mjuplong$ and a radius of $R_p = \genevaplanetradius\,\rjuplong$, residing in a close-in orbit with a period of \genevaperiodshort\,days. The host star is a $M_\star = \genevastarmass\,\msun$ A star rotating at $\vsini = \genevastarvsini\,\kms$, enabling us to characterize the spectroscopic transit of the brown dwarf via Doppler tomography. We find that \hatcurb{}, like other massive planets and brown dwarfs previously sampled, orbits in a low projected-obliquity orbit with $\lambda = \genevalambda \, ^\circ$. The low obliquities of these systems is surprising given all brown dwarf and massive planets with obliquities measured orbit stars hotter than the Kraft break. This trend is tentatively inconsistent with dynamically chaotic migration for systems with massive companions, though the stronger tidal influence of these companions makes it difficult to draw conclusions on the primordial obliquity distribution of this population. We also introduce a modeling scheme for planets around rapidly rotating stars, accounting for the influence of gravity darkening on the derived stellar and planetary parameters. 
\end{abstract}



\section{Introduction}
\label{sec:introduction}

Brown dwarf companions are a rarity around Sun-like stars \citep[e.g.][]{2000PASP..112..137M,2006ApJ...640.1051G,2011A&A...525A..95S}. These sub-stellar objects, with masses between 13 and 80\,\mjuplong{}, are seldom found within 0.1-0.2\,AU of a sun-like star \citep{2016AJ....151...85T}. As such, transiting examples of brown dwarf companions are particularly rare. Where available though, the transit geometry of a companion brown dwarf offers the same benefits as the planet counterparts for understanding their formation, evolution, and atmospheres. The densities of brown dwarfs may be related to their metallicity \citep[e.g.][]{2011ApJ...736...47B}, the obliquity angle offers a glimpse into their migrational history \citep{2009A&A...506..377T,2012ApJ...761..123S}, while their atmospheres may be compared to similarly irradiated giant planets \citep{2017AJ....154..242B}.  

The brown dwarf class encompasses the merging tail ends of giant planet formation and star formation. Core accretion may be responsible for massive giant planets and low mass brown dwarfs up to $\approx 40\,\mjuplong$ \citep{2012A&A...547A.112M}. Alternatively, disk instability may result in the formation of brown dwarfs within a similar mass range \citep[e.g.][]{2015MNRAS.452.1654N,2018ApJ...854..112M}. Hydrodynamic simulations from \citet{2009MNRAS.392..590B} can also reproduce a wide range of binaries with brown dwarf secondaries during the formation of star clusters. \citet{2014MNRAS.439.2781M} argue that the brown dwarf population can be divided along a gap at $35 \,\mjuplong < M_p < 55 \,\mjuplong$, where the eccentricity distributions of the two populations diverge. They argue that the lower mass end of the distribution may be the result of disk instability, while the higher mass brown dwarfs formed as a result of cloud fragmentation. \citet{2018ApJ...853...37S} further argues for a divide at $\sim 4\,\mjuplong$ based on metallicity trends in the planet occurrence rate: the less massive giant planets may have formed via core accretion, and therefore display a strong occurrence rate dependence with host star metallicity, while higher massed planets and brown dwarfs show no metallicity bias, a classical signature of disk instability. 

The properties of the orbiting brown dwarf may also depend on the mass of the host star. Radial velocity surveys have found massive planets to be more prevalent around massive stars \citep{2010PASP..122..905J,2014A&A...566A.113J,2018arXiv180909914B}, however massive planets can also spin up the host star, and lose orbital angular momentum in the process. Tidal dissipation via internal gravity waves in the convective-radiative boundary is efficient for G stars, and lacking in F type stars \citep{2010MNRAS.404.1849B,2014EAS....65..327G}. As such, massive planets may be more efficiently engulfed by stars below the Kraft break \citep{1967ApJ...150..551K,1970saac.book..385K}, and only surviving around higher mass stars. Though empirical studies making use of the hot-Jupiter population have often found that tidal in-spiral may take longer than expected for many systems, providing helpful constraints on the tidal dissipation coefficient across known planet-hosting stars \citep[e.g.][]{2012ApJ...751...96P,2018MNRAS.476.2542C}. The stellar binarity rate is also known to be an increasing function of stellar mass (both from observations, \citealt{2013ARA&A..51..269D}, and simulations, \citealt{2009MNRAS.392..590B}), and we may expect brown dwarf companions formed in the star formation process to be more abundant about higher mass stars.

We report the first transiting brown dwarf found around an A star. \hatcurb{} is a brown dwarf at the deuterium burning mass limit orbiting a $\teff = \genevastarteff$\,K A star with an orbital period of $P=\genevaperiodshort$ days. The transits were first identified by the HATSouth network \citep{2013PASP..125..154B}, and confirmed via photometric and spectroscopic follow-up observations that measured the radius and mass of the companion. Finally, blend scenarios are eliminated by measuring the Doppler tomographic transit of the brown dwarf, confirming that the transit and radial velocity signal are indeed originating from the A star, not a background binary. 

\section{Observations}
\label{sec:obs}

\subsection{Photometry}
\label{sec:photometry}

\begin{figure}
\plotone{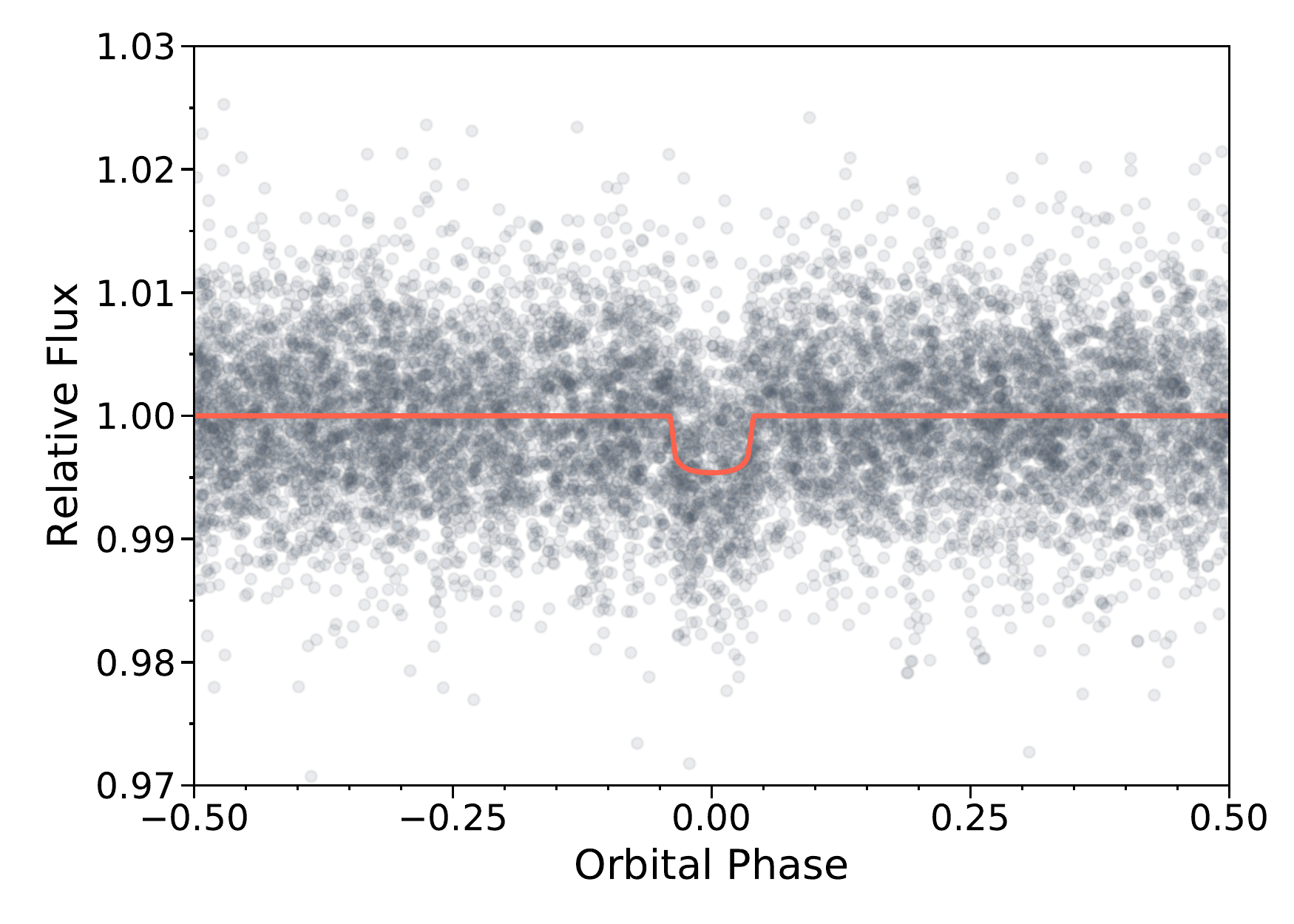}
\caption{
The unbinned $r$ band HATSouth discovery light curve of \hatcur{}. The light curve has been phase folded to the orbital period of $P=\genevaperiodshort$, and the best fit model, described in Section~\ref{sec:analysis}, is overlaid in red. 
\label{fig:hslc}}
\end{figure}

\begin{figure}
\plotone{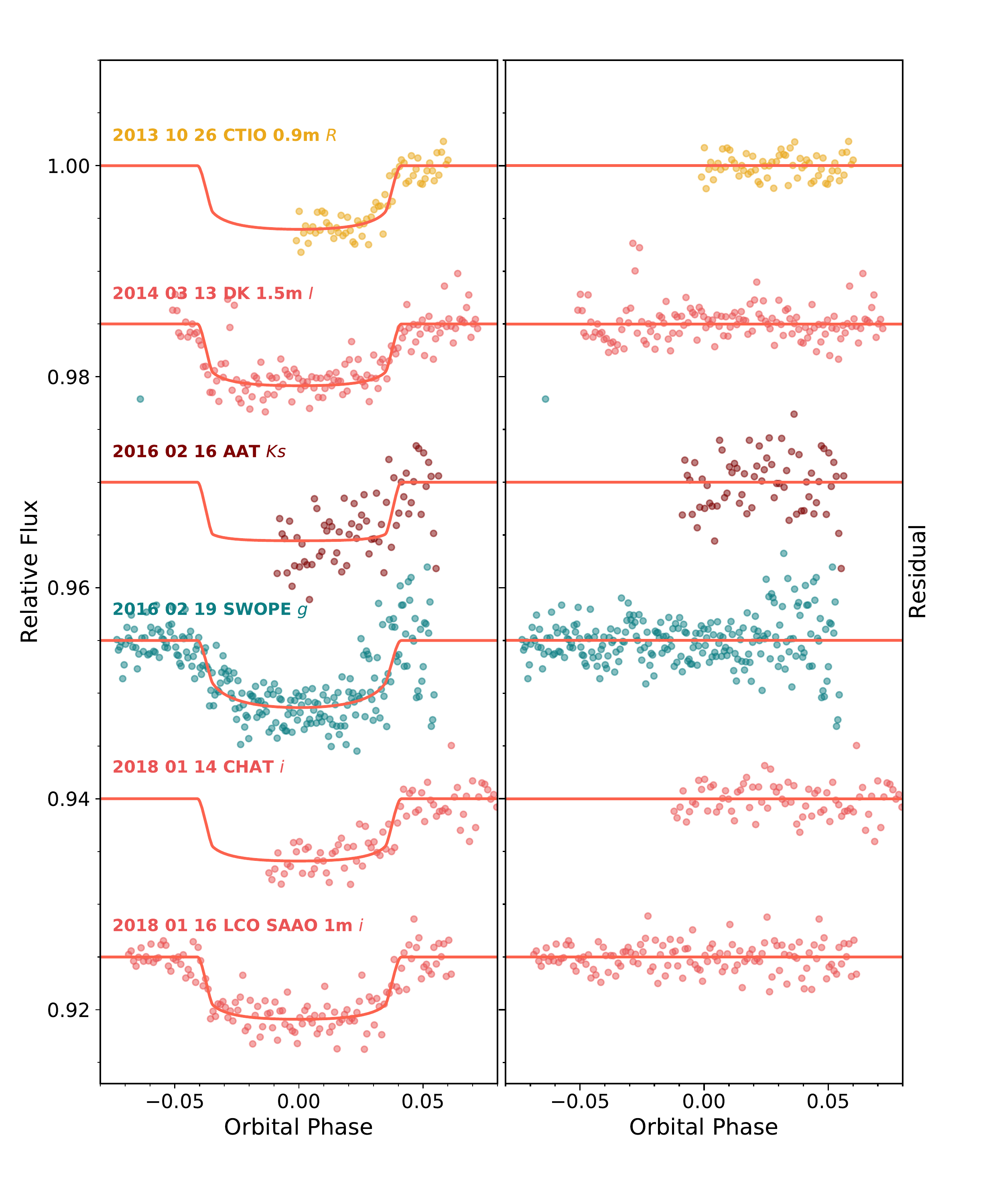}
\caption{
Follow-up light curves for the transit around \hatcur{}. The AAT IRIS2 $Ks$ band photometry have been binned to a phase of 0.001 $(\sim 3\,\mathrm{minutes})$, all other observations are presented unbinned. The best fit transit models are plotted in red over each set of observations. An arbitrary vertical offset has been applied to each observation for clarity. The residuals of each light curve after subtraction of the best fit model are shown in the right panel at the same scale. 
\label{fig:fulc}}
\end{figure}

The transits of \hatcurb{} were first identified by the HATSouth network \citep{2013PASP..125..154B}. To provide continuous coverage of large fields of the sky, HATSouth operates a network of telescopes across the Southern hemisphere, at Las Campanas Observatory in Chile, at the High Energy Spectroscopic
Survey (HESS) site in Namibia, and at Siding Spring Observatory (SSO) in Australia. The photometric reductions, including detrending via External Parameter Decorrelation \citep[EPD][]{2010ApJ...710.1724B} and Trend Filtering Algorithm \citep[TFA][]{2005MNRAS.356..557K}, and candidate finding via Box Least Squares \citep[BLS][]{2002A&A...391..369K} searches, are fully described in \citet{2013AJ....145....5P}. The HATSouth survey has discovered 60 planets to date, the full set of discoveries can be found at \url{https://hatsouth.org/}. The HATSouth discovery light curve for \hatcurb{} is shown in Figure~\ref{fig:hslc}, and the full light curve dataset is presented in Table~\ref{tab:lc_table} and Figure~\ref{fig:fulc}.

A series of photometric follow-up observations were obtained for the transits of \hatcurb{}, summarized in Table~\ref{tab:photobs}. These observations were gathered over the course of 5 years, covering photometric bands ranging from $g$ in the blue to $Ks$ in the near infrared. The follow-up transits help refine the ephemeris of the companion, and were used to search for color-dependence in the transit depth that may be indicative of astrophysical blending scenarios. 
The egress of \hatcurb{} was captured on 2013 Oct 26 with the 0.9\,m SMARTS Telescope at Cerro Tololo Inter-American Observatory (CTIO).  
A full $I$ band transit on 2014 Mar 13 was observed with the Danish Faint Object Spectrograph and Camera (DFOSC) on the Danish 1.54\,m (DK 1.54\,m) telescope at La Silla, Chile. 
A partial transit was obtained with the IRIS2 infrared camera on the Anglo-Australian Telescope at Siding Spring Observatory, Australia, on 2016 Feb 16. The observations were obtained in the $Ks$ band, and were reduced as per the procedure described in \citet{2014MNRAS.445.2746Z}. 
The 1\,m Swope telescope, located at Las Campanas Observatory, was used on 2016 Feb 19 to obtain a $g$ full transit using its e2v CCD camera. 
An $i$ band egress of \hatcurb{} was observed with the 0.7\,m Chilean-Hungarian Automated Telescope (CHAT) on 2018 Jan 14. CHAT is a dedicated transit-followup telescope located at Las Campanas Observatory, and makes use of a $2\mathrm{K}\times2\mathrm{K}$ back-illuminated CCD yielding a pixel scale of $0\arcsec.6\,\mathrm{pixel}^{-1}$ over a  field of view of $21\arcmin \times 21\arcmin$. Observations from the Las Cumbres Observatory \citep[LCO,][]{2013PASP..125.1031B} 1\,m telescope at the South African Astronomical Observatory (SAAO) on 2018 Jan 16 covered the full transit in $i$ band with the Sinistro camera. The data reductions were performed with the automated LCO pipeline, and photometric extraction was performed as per \citet{2015AJ....150...49B}.

\begin{deluxetable*}{llrrr}
\tablewidth{0pc}
\tabletypesize{\scriptsize}
\tablecaption{
    Summary of photometric observations
    \label{tab:photobs}
}
\tablehead{
    &
    &
    &
    &\\
    \multicolumn{1}{c}{Facility}          &
    \multicolumn{1}{c}{Date(s)}             &
    \multicolumn{1}{c}{Number of Images}\tablenotemark{a}      &
    \multicolumn{1}{c}{Cadence (s)}\tablenotemark{b}         &
    \multicolumn{1}{c}{Filter}            
}
\startdata
HS & 2011 Aug 24 -- 2012 Feb 14 & 9287 & 307 & $Rc$ \\
CTIO 0.9\,m & 2013 Oct 27 & 66 & 177 & $R$ \\
DK 1.5\,m & 2014 Mar 14 & 139 & 145 & $I$ \\
AAT 3.9\,m IRIS2 & 2016 Feb 16 & 1617 & 6 & $Ks$ \\
Swope 1\,m e2V & 2016 Feb 20 & 233 & 89 & $g$ \\
CHAT & 2018 Jan 15 & 90 & 200 & $i$ \\
LCO 1\,m SAAO Sinistro & 2018 Jan 16 & 131 & 163 & $i$ \\
\enddata 
\tablenotetext{a}{
  Outlying exposures have been discarded.
}
\tablenotetext{b}{
  Median time difference between points in the \lc. Uniform sampling was not possible due to visibility, weather, pauses.
}
\end{deluxetable*}

\begin{deluxetable*}{rrrrrrr}

\tablewidth{0pc}
\tabletypesize{\scriptsize}
\tablecaption{
        Differential photometry of \hatcur{}
    \label{tab:lc_table}
}
\tablehead{
    &
    &
    &
    & \\
    \multicolumn{1}{c}{BJD}          &
    \multicolumn{1}{c}{Mag (Raw)}\tablenotemark{a}             &
    \multicolumn{1}{c}{Mag (EPD)}            &
    \multicolumn{1}{c}{Mag (TFA)}            &
    \multicolumn{1}{c}{$\sigma$ Mag}      &
    \multicolumn{1}{c}{Instrument}         &
    \multicolumn{1}{c}{Filter}            
}
\startdata
2455797.9170947 & 13.02009 & 12.73791 & 12.67781 & 0.00464& HATSouth & Rc\\
2455798.9199963 & 13.02911 & 12.73387 & 12.67026 & 0.00485& HATSouth & Rc\\
2455799.6748836 & 12.96931 & 12.73267 & 12.67586 & 0.00447& HATSouth & Rc\\
2455800.6715830 & 12.97502 & 12.72716 & 12.67409 & 0.00426& HATSouth & Rc\\
2455801.6742945 & 12.96676 & 12.71739 & 12.6602 & 0.00458& HATSouth & Rc\\
\enddata 
\tablenotetext{a}{
This table is available in a machine-readable form in the online journal. A portion is shown here for guidance regarding its form and content.\\
Raw, EPD, and TFA magnitudes are presented for HATSouth light curves. The detrending and potential blending may cause the HATSouth transit to be shallower than the true transit in the EPD and TFA light curves. This is accounted for in the global modeling by the inclusion of a dilution factor. Follow-up light curves have been treated with EPD simultaneous to the transit fitting. Pre-EPD magnitudes are presented for the follow-up light curves.
}

\end{deluxetable*}

\subsection{Spectroscopy}
\label{sec:spectroscopy}

A series of spectroscopic observations of \hatcur{} were obtained to measure the mass of the companion, constrain the properties of the host star, and to measure the line profile variations during the transit of \hatcurb{}. These observations are summarized in Table~\ref{tab:specobssummary}.

\begin{figure}
\plotone{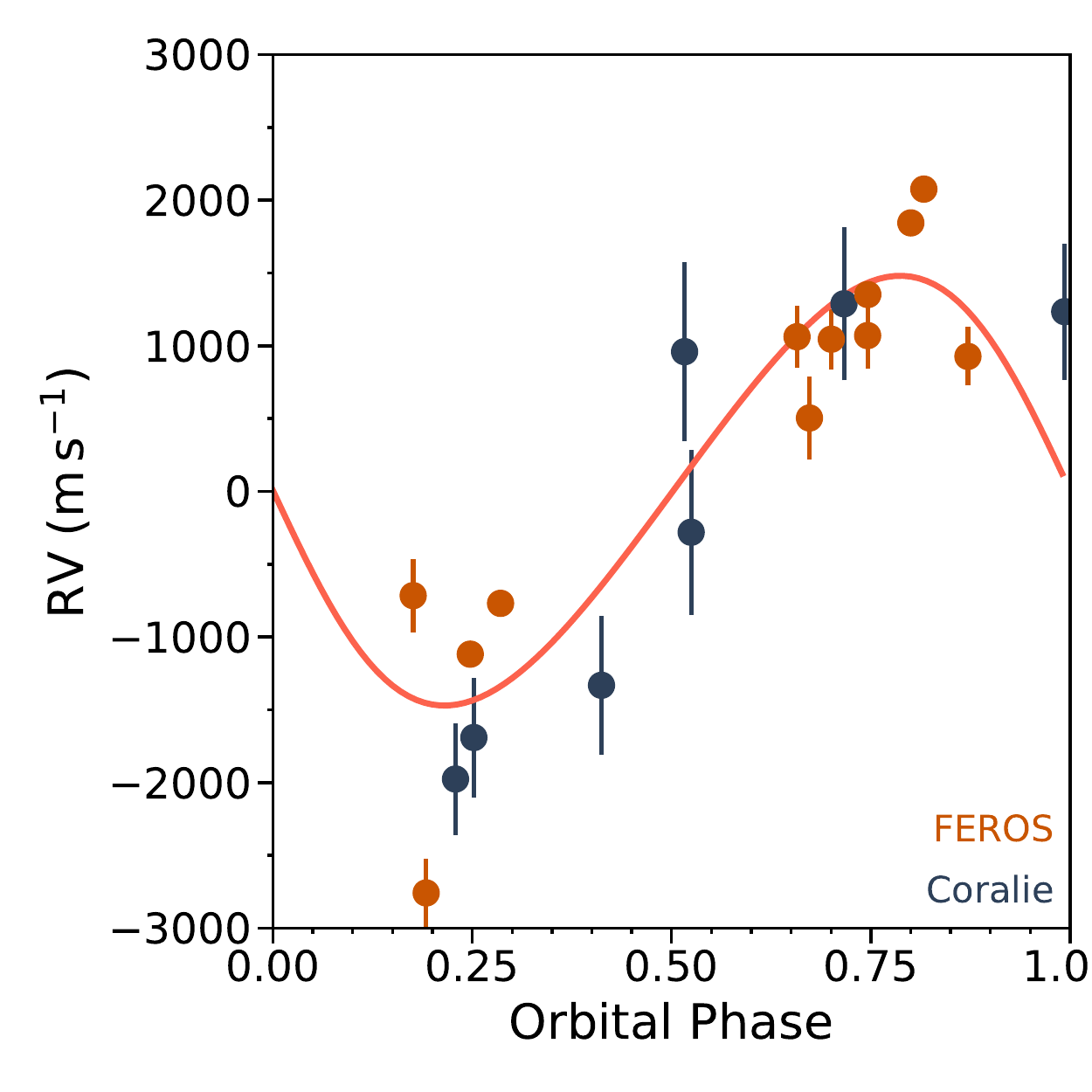}
\caption{
The radial velocities of \hatcur{}, showing the \genevaKrv\,\kms orbit induced by the orbiting brown dwarf. Velocities from CORALIE are plotted in blue, FEROS in orange. The systemic velocities of each instrument have been subtracted. The best fit model orbit is plotted in red.  
\label{fig:rv}}
\end{figure}

Seven observations were obtained with the CORALIE spectrograph \citep{2001Msngr.105....1Q} on the Euler 1.2\,m telescope at La Silla Observatory, Chile. Spectra from CORALIE covers the range of $3900-6800$\,\AA{} at a resolution of $\lambda / \Delta \lambda \equiv R = 60000$. Our observations were obtained with integration times of $1800-3600$\,s, yielding a signal-to-noise ratio of $\sim 20$ per resolution element. An additional twelve observations were obtained with FEROS \citep{1998SPIE.3355..844K} on the MPG 2.2\,m telescope at La Silla. FEROS is a fiber fed spectrograph with spectral resolution of $R = 48000$ over the wavelength range of $3500-9200$\,\AA. The observations were of 1800\,s in exposure time, yielding a signal-to-noise of $\sim 80$ at peak. Spectroscopic reductions and radial velocity measurements for the CORALIE and FEROS observations were performed using the CERES pipeline \citep{2017PASP..129c4002B}. The spectra are extracted via the optimal extraction technique \citep{1986PASP...98..609H,1989PASP..101.1032M}, with weights determined from the Quartz-lamp illuminated calibrations, while the wavelength calibration is obtained via a 2D Chebyshev polynomial fit to the ThAr arc lamp lines. Radial velocities are derive from the spectra via cross correlations against a binary mask of a G2 star, similar to the procedure described in \citep{2003Msngr.114...20M}. The radial velocities from CORALIE and FEROS are listed in Table~\ref{tab:rv_table}. The radial velocity orbit is plotted in Figure~\ref{fig:rv}.

Spectroscopic observations during transit can reveal the passing shadow of the transiting companion as parts of the rotating stellar disk are occulted. These time-series observations record the stellar line profile variations due to the occultation, and have been widely used to measure the projected obliquity angle of planets \citep{1924ApJ....60...15R,1924ApJ....60...22M,2000A&A...359L..13Q}, as well as being a final confirmation for the nature of the transiting companion \citep{2010MNRAS.407..507C}. 

To measure the spectroscopic transit of \hatcurb{}, we obtained a series of observations with the Magellan Inamori Kyocera Echelle \citep[MIKE,][]{2003SPIE.4841.1694B} on the 6.5\,m Magellan Clay telescope at Las Campanas Observatory on 2017 Dec 27. The observations span from 00:46 to 07:45 UTC, covering the full transit, and were obtained over the airmass range of 2.1--Zenith--1.1 over the course of the night, with the seeing remaining below $1\arcsec$ throughout. The observations were obtained with the 0.\arcsec35 slit, yielding the highest possible resolution for the spectrograph of $R=85000$ in the blue camera ($3200-5000$\,\AA), and $R=65000$ in the red ($4900-10000$\,\AA). An integration time of 900\,s was adopted for both the blue and red arms, with ThAr arc lamp exposures every 30 minutes providing the wavelength solution. Spectral extraction was performed with the Carnegie \emph{Carpy} package \citep{2000ApJ...531..159K,2003PASP..115..688K}. 

\begin{figure}
\plotone{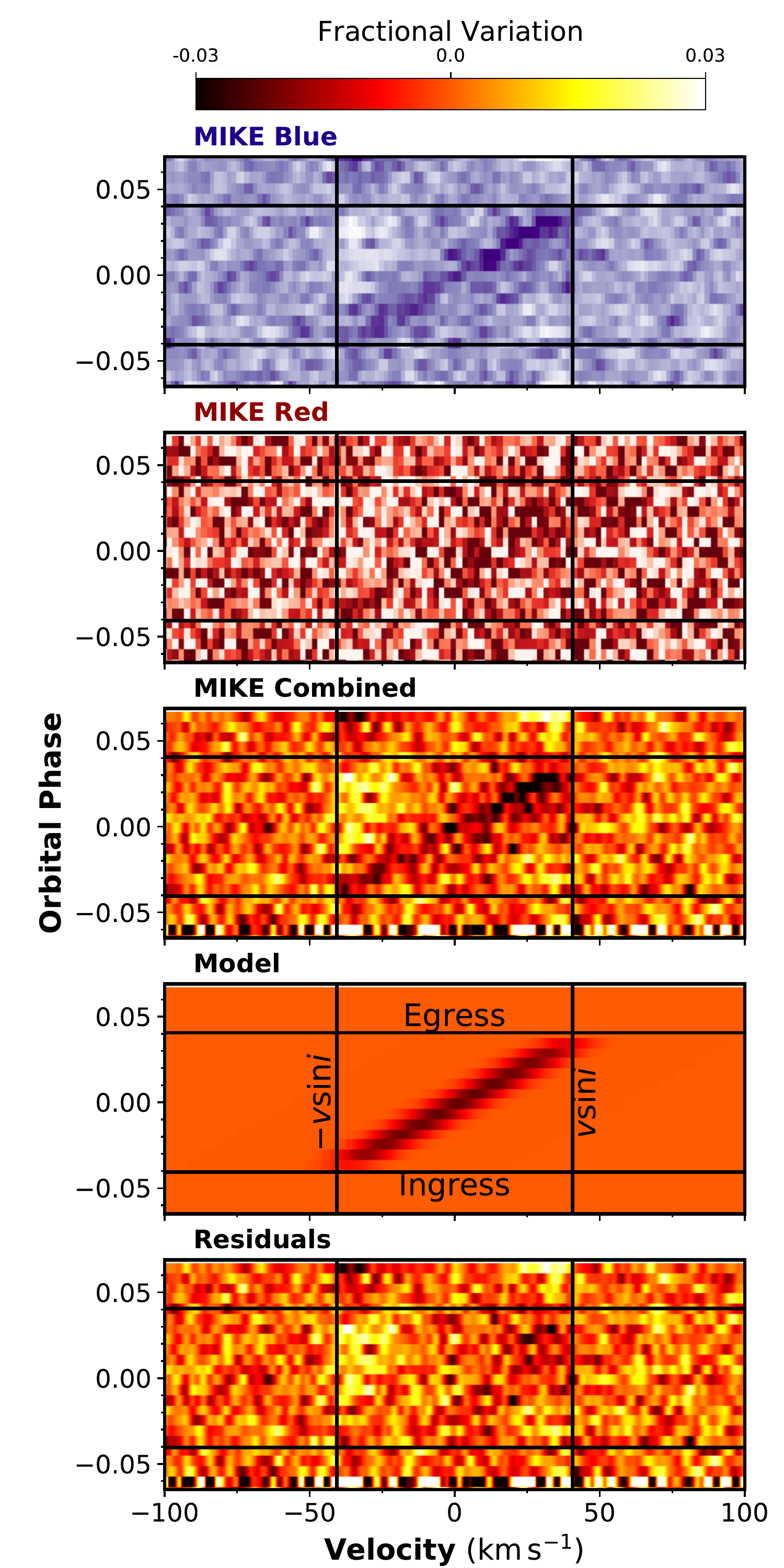}
\caption{
The Doppler tomographic transit of \hatcurb{} as measured by Magellan / MIKE on 2017 Dec 27. The line profile residuals are plotted as a function of orbital phase and velocity. The top two panels show the transit as seen with the blue and red cameras. The middle panel shows the combined dataset. The bottom two panels show the best fit model and the residuals after model subtraction. 
\label{fig:dt}}
\end{figure}

The Doppler tomographic analysis of the MIKE observations follow the process described in \citet{2018arXiv180700024Z}. The line profiles are derived using a least-squares deconvolution \citep{1997MNRAS.291..658D} against a set of synthetic non-rotating spectral templates generated using \emph{SPECTRUM} \citep{1994AJ....107..742G} with ATLAS9 model atmospheres \citep{Castelli:2004}. The line profiles are derived from order to order, and weighted averaged based on the signal-to-noise of the line profile peak height. The line profile variations, plotted as a function of time, are shown in Figure~\ref{fig:dt}. The transit signal of the brown dwarf is clearly seen as the dark trail extending from bottom left (ingress) to top right (egress). 

We also attempted a Doppler tomographic observation with the echelle spectrograph on the 2.5\,m Ir\'{e}n\'{e}e du Pont telescope at Las Campanas Observatory on 2016 Feb 20. A total of 22 observations were obtained, and line profiles were measured using the least-squares deconvolution analysis, but were of too low signal-to-noise to reveal the shallow planetary shadow.

\begin{deluxetable*}{llrrrrr}
\tablewidth{0pc}
\tabletypesize{\scriptsize}
\tablecaption{
    Summary of spectroscopic observations\label{tab:specobssummary}
}
\tablehead{
\\
    \multicolumn{1}{c}{Telescope/Instrument} &
    \multicolumn{1}{c}{Date Range}          &
    \multicolumn{1}{c}{Number of Observations} &
    \multicolumn{1}{c}{Resolution}          &
    \multicolumn{1}{c}{Observing Mode}          
}
\startdata
Euler 1.2\,m CORALIE & 2013 Nov 21 -- 2014 Sep 12 & 7 & 60000 & RV\\
MPG 2.2\,m FEROS & 2013 Dec 26 -- 2015 Feb 03 & 12 & 48000 & RV\\
du Pont 2.5\,m echelle & 2016 Feb 20 & 22 & 45000 & Transit \\
Magellan 6.5\,m MIKE-blue & 2017 Dec 27 & 24 & 85000 & Transit \\
Magellan 6.5\,m MIKE-red & 2017 Dec 27 & 24 & 65000 & Transit \\
\enddata 

\end{deluxetable*}

\begin{deluxetable}{rrrl}
\tablewidth{0pc}
\tabletypesize{\scriptsize}
\tablecaption{
       Relative radial velocities of \hatcur{}
    \label{tab:rv_table}
}
\tablehead{ \\
    \multicolumn{1}{c}{BJD}          &
    \multicolumn{1}{c}{RV}\tablenotemark{a}             &
    \multicolumn{1}{c}{$\sigma$ RV}      &
    \multicolumn{1}{c}{Instrument}      \\
    \multicolumn{1}{c}{(UTC)} &
    \multicolumn{1}{c}{$(\mathrm{m\,s}^{-1})$} &
    \multicolumn{1}{c}{$(\mathrm{m\,s}^{-1})$} & 
}
\startdata
2456617.7421542602 & 34249 & 713 & CORALIE\\
2456618.6623599301 & 37513 & 979 & CORALIE\\
2456619.6738256998 & 34534 & 765 & CORALIE\\
2456620.6916636401 & 43412 & 2224 & CORALIE\\
2456727.6056944099 & 34893 & 887 & CORALIE\\
2456730.5911809001 & 37459 & 868 & CORALIE\\
2456731.5789716602 & 37184 & 1144 & CORALIE\\
2456912.8653775398 & 35944 & 1055 & CORALIE\\
2456652.7070320700 & 37107 & 61 & FEROS\\
2456654.7279559700 & 37831 & 32 & FEROS\\
2456704.6360209300 & 34637 & 34 & FEROS\\
2456705.6795871798 & 37599 & 35 & FEROS\\
2456706.5959924399 & 34986 & 38 & FEROS\\
2457030.5916852201 & 36682 & 212 & FEROS\\
2457035.8515796200 & 36817 & 223 & FEROS\\
2457037.8210289800 & 36800 & 220 & FEROS\\
2457053.8246719800 & 35039 & 265 & FEROS\\
2457054.7633720501 & 36259 & 300 & FEROS\\
2457055.7437583399 & 32997 & 247 & FEROS\\
2457056.7893937798 & 36825 & 241 & FEROS\\
\enddata 
\tablenotetext{a}{
 Internal errors excluding the component of astrophysical/instrumental jitter considered in Section 3. 
}

\end{deluxetable}

\subsection{Lucky imaging}
\label{sec:luckyimaging}

\begin{figure*}
\plottwo{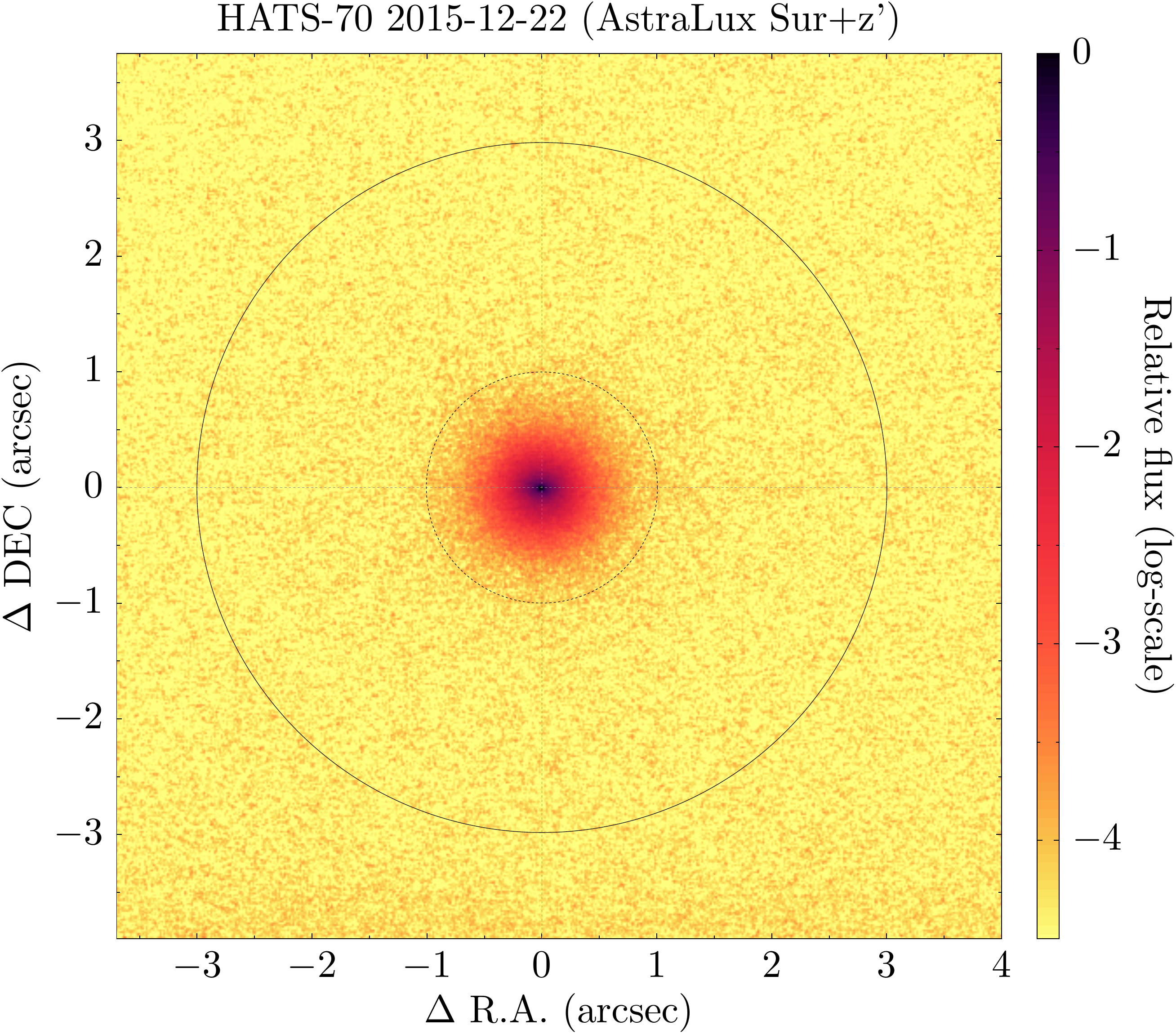}{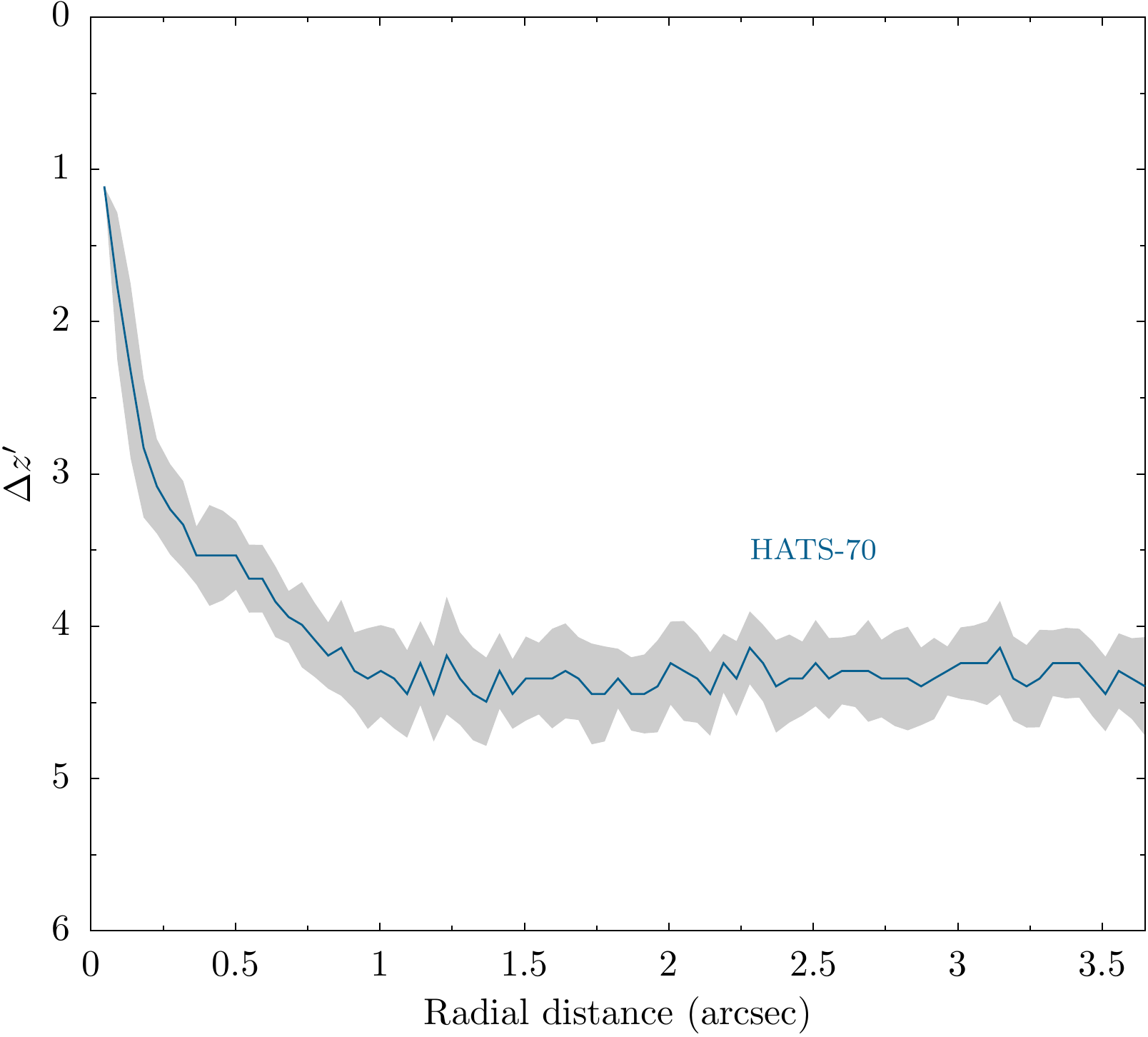}
\caption{
\textbf{Left:} AstraLux Sur $z'$-band image of \hatcur{}, showing no stellar companions detected. \textbf{Right:} The contrast curve from the Astralux Sur observation, with a  $\mathrm{FHWM}_\mathrm{eff} = 59.5\pm4.8\,\mathrm{mas}$. The shaded region represents the uncertainty on the contrast curve given by the scatter  along the azimuthal direction at a given radial distance. 
\label{fig:luckyimage}}
\end{figure*}

To further check for nearby stellar companions to \hatcur, we obtained $z'$ band lucky imaging observations with the Astralux Sur camera \citep{2009Msngr.137...14H} on the 3.58\,m New Technology Telescope at La Silla Observatory. The observations and reductions follow the procedure in \citet{2016AJ....152..108E}. By combining 10\% of the images, we obtain an effective point-spread function with an effective full width at half maximum of $\mathrm{FHWM}_\mathrm{eff} = 3.92\pm0.32$ pixels, or $59.6\pm4.8\,\mathrm{mas}$ at a pixel scale of $15.20 \mathrm{mas\,pixel}^{-1}$ \citep{2017A&A...599A..70J}. We did not detect any stellar companions within $2\arcsec$ for \hatcur{}. The resulting Astralux images and contrast curves are show in Figure~\ref{fig:luckyimage}

\section{Analysis}
\label{sec:analysis}

\subsection{Properties of the host star}
\label{sec:hoststar}

\begin{figure*}
\plotone{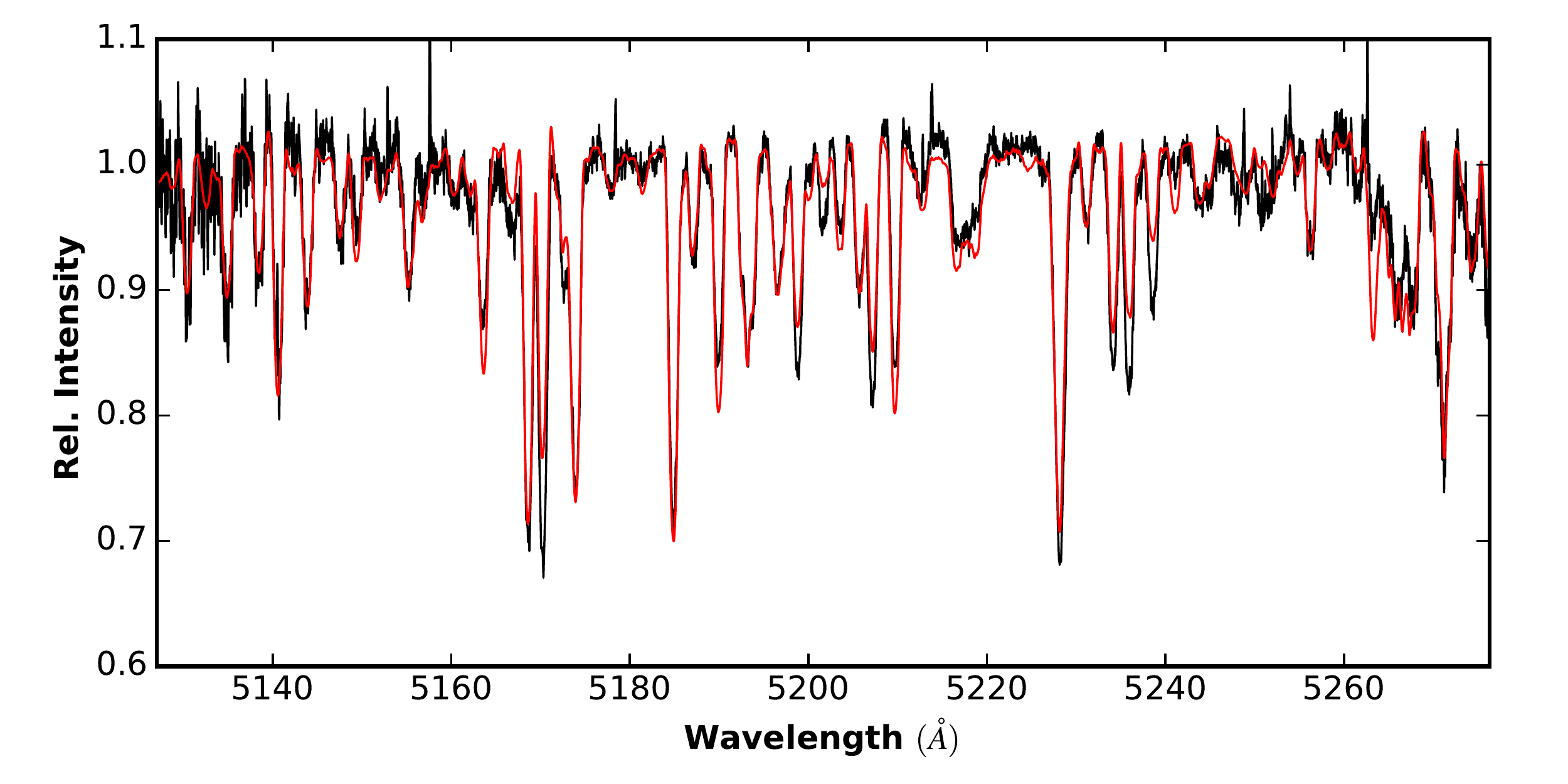}
\caption{
The out-of-transit averaged MIKE spectrum of \hatcur{} over the Mg b line order. The observed spectrum is plotted in black, a synthetic template of $\teff=7750$\,K, $\logg = 4.0$\,dex, $\feh = 0.0$\,dex is plotted in red. The synthetic template has been convolved against the least-squares deconvolution-derived line broadening kernel to match the observation.  
\label{fig:spectrum}}
\end{figure*}

With a broadband color of $J-K=0.099$ for \hatcur{}, it was evident that the host star is of an early spectral class. An initial estimate of the effective temperature with the $J-K$ color \citep{2006AJ....131.1163S} and Gaia parallax \citep{2018A&A...616A...1G} confirms that the target star is indeed an A dwarf. We make use of the combined out-of-transit MIKE spectra for a spectral match against a library of synthetic spectra generated with SPECTRUM (see Section~\ref{sec:spectroscopy}). The template spectra are broadened by the observed broadening kernel derived from a least-squares deconvolution against a non-rotating stellar template. This convolution matches the rotational and macroturbulent broadening of the templates to the observation perfectly, as well as accounting for the radial velocity shift of the target star. We perform a least squares fit of the spectrum over $5150-5350$\,\AA{} against a grid of synthetic templates from \citet{2012Natur.486..375B}, at step sizes of 500\,K in \teff{} and 0.5\,dex in \feh. We adopt a surface gravity of $\log g = 4.0$ for this analysis. The $\log g$ will be eventually refined via transit-derived stellar densities within the global analysis (Section~\ref{sec:globalmodel}). We find a best fitting template of $T_\mathrm{eff} = 7750\,\mathrm{K}$ and $\mathrm{[Fe/H]} = 0.0$\,dex, confirming the target star is indeed a near-solar metallicity A-star. The best fit spectral template is plotted against the observed spectrum in Figure~\ref{fig:spectrum}. We note, however, that without the lack of a substantial set of standard star comparisons obtained with MIKE in a similar setup, we cannot accurately estimate the uncertainties involved in our stellar parameter estimates. The final stellar parameters are provided within the global analysis via an spectral energy distribution (SED) fit (Appendix~\ref{sec:gravdark_SED}). 

Stars hotter than the Kraft break have radiative envelopes, and lack the mass-loss process of solar-type stars that spin down over their main sequence lifetimes. As such, A stars are often found rotating at $10-100\,\kms$. This rapid rotation hinders precise radial velocity measurements, but enables us to obtain a Doppler tomographic measurement of the planetary transit. An accurate estimation of the rotational velocity of the host star is essential in determining the projected obliquity angle from this observation. As such, we model the out-of-transit line broadening profiles derived from the least-squares deconvolution process to measure the \vsini{} of \hatcur{}. The line profiles are modeled with a rotational, macroturbulent, and instrumental broadening kernel as per \citet{2018arXiv180700024Z}. The contributions of rotational and macroturbulent broadening are computed via a disk integration as per \citet{2005oasp.book.....G}. The model profile is then convolved against a Gaussian of width 4.6\,\kms{} to account for the instrumental broadening of MIKE's red arm. We fit for values of rotational and macroturbulent broadening via a Markov chain Monte Carlo analysis using the ensemble sampler \emph{emcee} \citep{2013PASP..125..306F}, and find a best fit value of $\vsini = 40.58\pm0.34\,\kms$ and $v_\mathrm{macro} = 6.0\pm 1.2\,\kms$. Since it is difficult to measure accurate uncertainties from the LSD profiles, we measure the velocity broadening values for each observed spectrum independently, and take the scatter in the values as our uncertainty. To better understand the instrument dependence of our broadening velocity measurements, we perform the same analysis on spectra from the other facilities. We measure $\vsini = 40.16 \pm 0.51\,\kms$ and $v_\mathrm{macro} = 5.80\pm 0.23\,\kms$ from the Magellan MIKE blue arm, $\vsini = 40.29 \pm 0.25 \,\kms$ and $v_\mathrm{macro} = 5.85 \pm 0.26\,\kms$ from FEROS, and $\vsini = 40.14 \pm 0.66 \,\kms$ and $v_\mathrm{macro} = 5.98 \pm 0.35\,\kms$ from the du Pont echelle. The broadening velocity measurements are consistent to $1\sigma$ with the Magellan red arm measurement. We adopt the Magellan red arm $\vsini$ and $v_\mathrm{macro}$ velocities as priors for the global modeling process in Section~\ref{sec:globalmodel}. 

\subsection{Search for light curve modulation and additional companions}

We search for the additional signals in the HATSouth discovery light curves indicative of rotational modulation, pulsations, or additional transiting companions. Running a generalized Lomb-Scargle analysis \citep{2009A&A...496..577Z} on the transit-removed HATSouth light curve yields no significant detection. The highest peak in the periodogram has a false alarm probability of 50\%, with a 95\% confidence upper limit on the semi-amplitude of 0.9\,mmag. To search for additional companions, we run another iteration of BLS on the transit-removed light curves, revealing no additional transit signals present. The highest peak in the BLS spectrum has a period of 0.2622\,days and a signal-to-noise ratio of 7, below our threshold for identifying significant detections in HATSouth light curves. We note that \hatcur sits at the edge of the instability strip for $\delta$ scuti stars, though unlike many previous A-star planet hosts \citep[e.g.][]{2010MNRAS.407..507C,2017MNRAS.471.2743T}, it lacks signatures of strong pulsations in the HATSouth discovery light curves, follow-up light curves, or in the Doppler tomographic line profile residuals.

\subsection{Global modeling of system parameters}
\label{sec:globalmodel}

We perform a global modeling of the system parameters to determine the final stellar and brown dwarf properties. The modeling of \hatcurb{} is complicated by the rapid rotation of the host star and the non-negligible mass of the secondary. In particular, the rapid rotation of the host star 1) makes it difficult to determine precise stellar parameters from spectra, due to the broadening and blending of key lines, 2) requires the effects associated with stellar gravity darkening to be incorporated in the modeling of the system, accounting for stellar oblateness and the latitudinal-dependence in the illumination of the stellar disk. 

To account for these effects, we make use of the gravity darkening code made available by \citet{2018AJ....155...13H} in modeling the transit light curves. The stellar mass and radius are estimated by fitting the SED simultaneous to the global modeling. The Geneva isochrones \citep{2012A&A...537A.146E} are employed to constrain the stellar oblateness, in order to estimate the magnitude of the gravity darkening effect, and to properly account for the geometric dependence between the apparent luminosity and the line of sight inclination of the star. A full description of the treatment of gravity darkening in the SED modeling can be found in Appendix~\ref{sec:gravdark_SED}. 

We make use of the discovery light curves, follow-up light curves, radial velocities from CORALIE and FEROS, the Doppler tomographic observation from Magellan/MIKE, photometric magnitudes from APASS \citep{2016yCat.2336....0H} and 2MASS \citep{2006AJ....131.1163S}, and Gaia DR2 \citep{2018A&A...616A...1G} parallax information in our analysis. Free parameters include the orbital period $P$, transit centroid $T_c$, radius ratio $R_p/R_\star$, and transit inclination $i$. The stellar and companion properties are fitted for directly, with free parameters for the mass $M_\star$, radius $R_\star$, and metallicity [Fe/H] of the host star, and mass of the brown dwarf companion $M_p$. The radial velocities are computed based on the tested masses at each step, along with the eccentricity free parameters $\sqrt{e} \cos \omega$ and $\sqrt{e} \sin \omega$; we also account for jitter as per \citet{2016MNRAS.457.3637H}, and allow a systemic offset for velocities from each facility. The projected orbital obliquity $\lambda$, rotational broadening velocity $v\sin I_\star$, and stellar macroturbulence $v_\mathrm{macro}$ are used to constrain the Doppler tomographic signal. We allow for additional parameters to describe the SED, including line of sight stellar inclination $I_*$, parallax, and interstellar reddening $E(B-V)$. Correlated noise in the follow-up light curves are corrected for with linear detrending coefficients against parameters including time, airmass, target star position on the detector, and full-width at half-maximum of the point-spread-function. The transits in the discovery light curve are often shallower than those in the follow-up light curves due to various detrending processes during the initial signal search. We account for this by multiplying the transit model with a dilution factor to match the HATSouth light curves. Photometric uncertainties can often be underestimated, we therefore inflate the per-point uncertainties of each of the discovery and follow-up light curves such that the reduced $\chi^2$ is at unity before fitting the datasets.

In the global modeling, the gravity darkening exponent and the quadratic limb darkening parameters are fixed to their interpolated values from \citet{Claret:2011}, and are listed in Table~\ref{tab:planetparam}. To model the gravity darkening effect in both the light curve and the SED, we interpolate the Geneva isochrones for a stellar oblateness value $(R_\mathrm{pole}/R_\mathrm{eq})$, given the stellar mass, radius, and equatorial rotational velocity tested at each iteration of the minimization. The resulting posterior for the stellar oblateness is reported in Table~\ref{tab:stellar}.    

The Doppler tomographic transit is modeled via a 2D integration of the stellar surface occulted by the planet, to include the effects of differential limb darkening, instrument broadening, and macroturbulent broadening of the planet's shadow \citep[as per][]{2017AJ....153..211Z}. In addition, to account for the broadening induced by the long integration times of our spectra, we integrate over three separate epochs for each spectral observation. We assume a quadratic limb darkening coefficient at the Kepler band from \citet{Claret:2011} when computing the line profiles for the Doppler tomographic model. The MIKE blue and red arm datasets are modeled separately as to account for their respective instrument resolutions. 

The posterior is sampled with a Markov chain Monte Carlo analysis, using the affine invariant ensemble sampler \emph{emcee} \citep{2013PASP..125..306F}. The rotational and macroturbulent broadening velocities are constrained by Gaussian priors about their spectroscopically measured values. The stellar metallicity is constrained by a Gaussian prior about the Galactic disk metallicity at 0.15-1.00 Gyr \citep{2003A&A...409..523R}. The system parallax is constrained by a prior about the Gaia DR2 value \citep{2018A&A...616A...1G}, with the systematic correction from \citet{2018ApJ...862...61S} applied. Reddening of the SED is also constrained to be below the local maximum from \citet{2011ApJ...737..103S}. The inclination $I_\star$ of the rotation axis is constrained by a $\cos I_\star$ prior. 

We also make the same analysis with the Dartmouth isochrones \citep{2008ApJS..178...89D} for a comparison against models not including rotation. Similar to the analysis with the Geneva rotational isochrones, we make use of the pre-computed magnitudes of the Dartmouth isochrones to perform a simultaneous SED fit with the system modeling. In addition to the APASS and 2MASS magnitudes, we also make use of the $G$, $BP$, and $RP$ magnitudes from Gaia for the SED fitting. These Gaia magnitudes are not used for the Geneva isochrone modeling since there is no available limb darkening coefficients in the Gaia bands, making it difficult for us to compute magnitudes from the disk-integrated SEDs. 

The resulting stellar and planetary parameters are displayed in Tables~\ref{tab:stellar} and \ref{tab:planetparam} respectively. Where appropriate, the priors have been indicated in the tables. 

\begin{figure}
\plotone{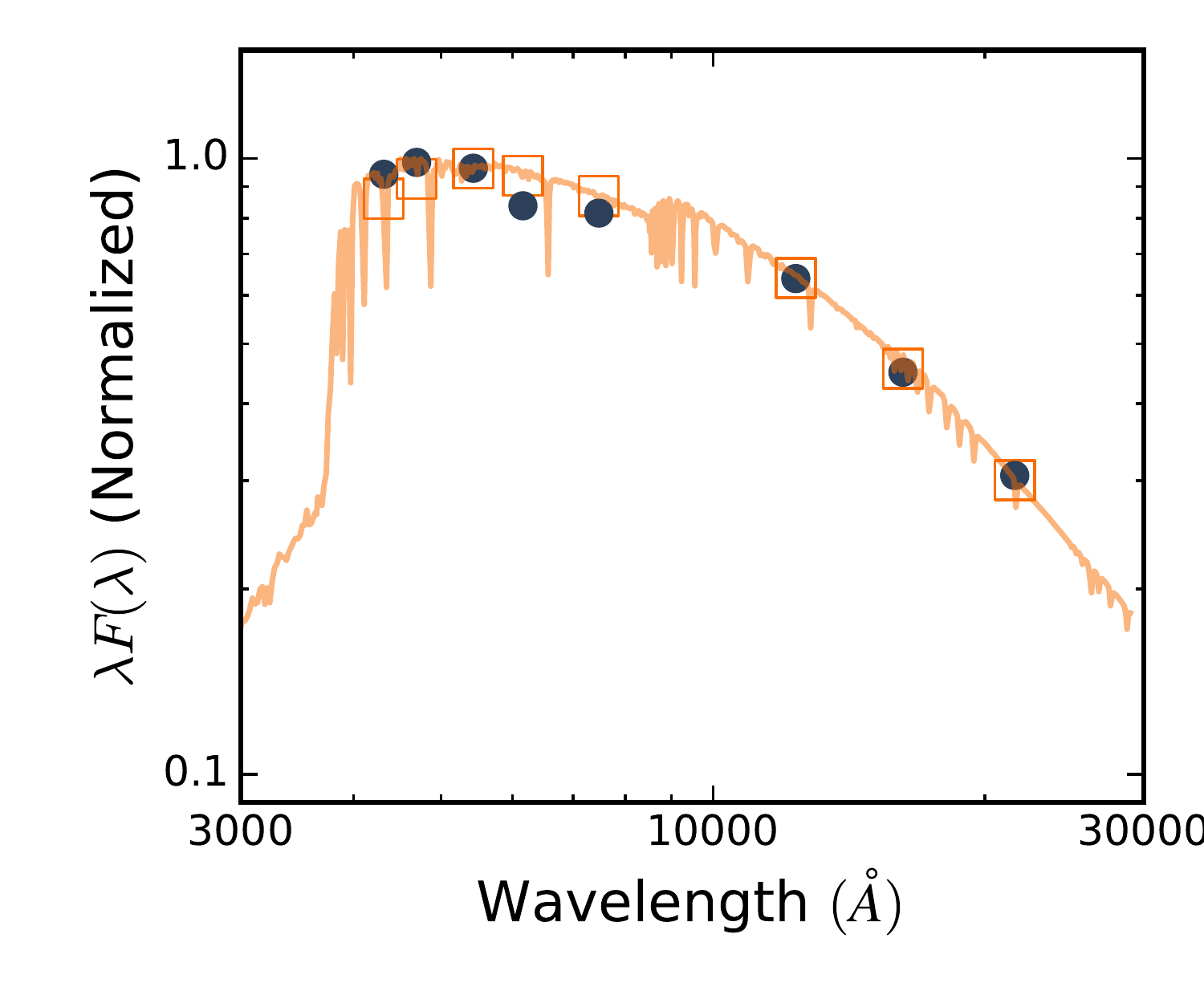}\\
\plotone{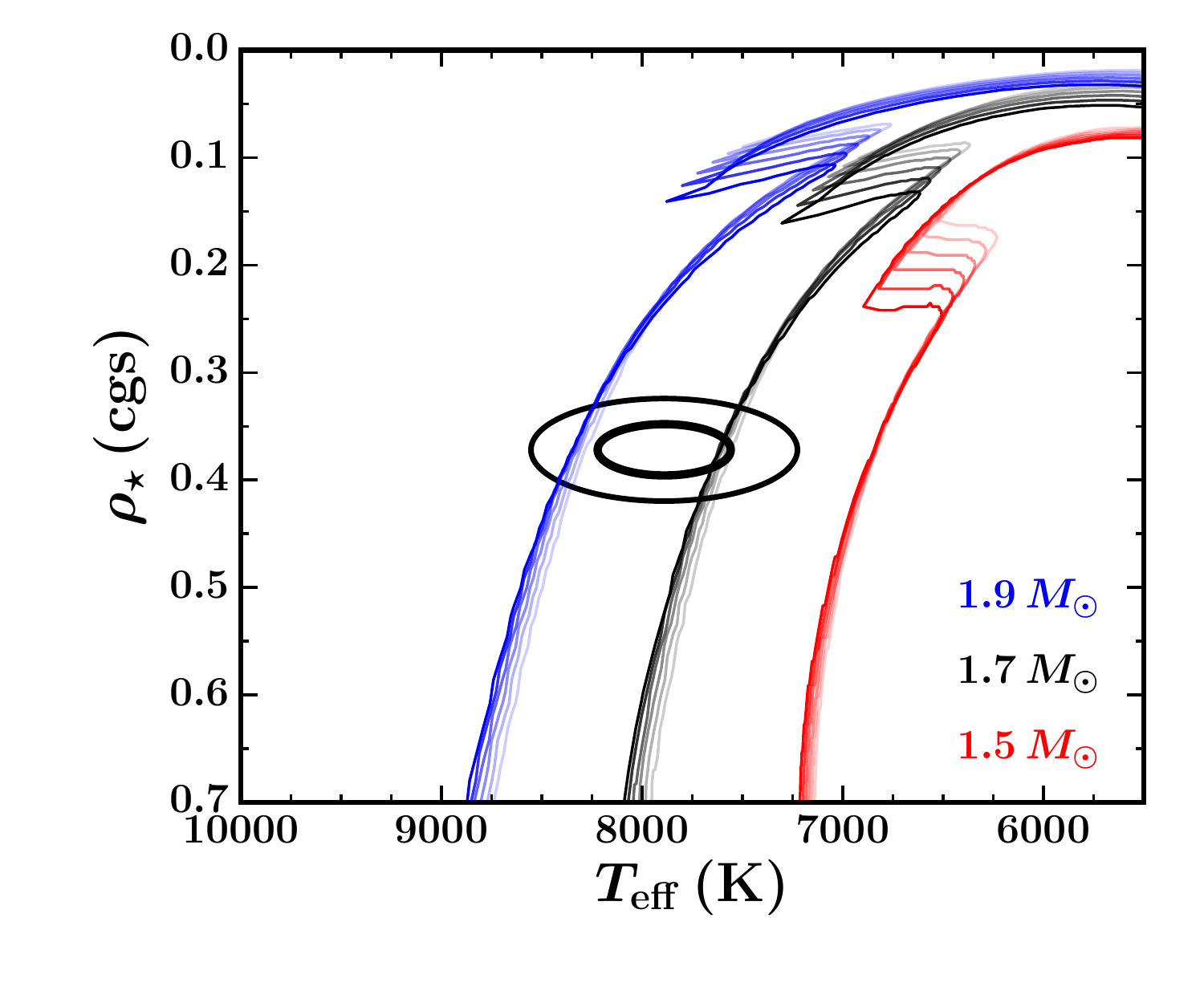}
\caption{
\textbf{Top}: The SED of \hatcur{}. Optical magnitudes for the $B$, $V$, $g$, $r$, and $i$ bands are from the APASS \citep{2016yCat.2336....0H} survey, the $J$, $H$, and $K$ band magnitudes from 2MASS \citep{2006AJ....131.1163S}. The model SED is computed by a disk integration over a gravity darkened model stellar surface with ATLAS9 models \citep{Castelli:2004}, and accounts for the best fit inclination of the stellar rotation axis.  \textbf{Bottom}: Error ellipses showing the SED-inferred effective temperature of \hatcur{} against the transit-derived stellar density $\rho_\star$. The 1 and $2\sigma$ error ellipses are marked. Geneva evolution tracks \citep{2012A&A...537A.146E} are plotted in the background for stars of mass 1.5, 1.7, and 1.9\,\msun{}. The gradient of the lines indicate the modeled stellar rotation of each track, with no rotation ($\Omega / \Omega_c = 0$) being darkest, $\Omega / \Omega_c = 0.5$ lightest. 
\label{fig:sed}}
\end{figure}

We find \hatcurb{} to be a $M_p = \genevaplanetmass \,\mjuplong$, $R_p = \genevaplanetradius \,\rjuplong$ brown dwarf orbiting a $M_\star = \genevastarmass\,\msun$ A star. The brown dwarf is orbiting well aligned with the spin of its host star, with a projected obliquity of $|\lambda| = \genevalambda \,^\circ$. The results from the Geneva rotational isochrone models are consistent with those from the non-rotating Dartmouth isochrone modeling. Figure~\ref{fig:sed} shows the best fit Geneva-based SED to the APASS and 2MASS magnitudes, as well as the transit-inferred stellar density against the SED-inferred effective temperature along the Geneva evolution tracks. 

To test the robustness of our analysis, we also model the system with the orbital eccentricity fixed to zero, but found no significant differences in $M_p$ and $R_p$ compared to the results presented here. To assess the levels of correlated noise remaining in the follow-up light curves following the simultaneous detrending, we follow equation 2 of \citet{2008ApJ...683.1076W} and bin the light curve residuals into consecutively larger segments, and compare the scatter of the binned residuals against that of the unbinned light curve. We find that the $\beta$ coefficient is below 1.3 for each of the follow-up light curves post-detrending; for reference uncorrelated noise should result in $\beta$ coefficients of 1.

\begin{deluxetable*}{lrrr}
\tablewidth{0pc}
\tabletypesize{\scriptsize}
\tablecaption{
    Stellar parameters for \hatcur{}
    \label{tab:stellar}
}
\tablehead{ \\
    \multicolumn{1}{c}{~~~~~~~~Parameter~~~~~~~~}   &
    \multicolumn{1}{c}{\textbf{Rotational model}} &
    \multicolumn{1}{c}{Non-rotational model} &
    \multicolumn{1}{c}{Prior\tablenotemark{a}} 
}
\startdata
\sidehead{Catalogue Information}
~~~~Tycho-2 \dotfill & 7103-114-1 && \\
~~~~2MASS \dotfill& J07162509-3114397 && \\
~~~~Gaia DR2 \dotfill & 5605119586158973440 && \\
~~~~Gaia RA (J2000) \dotfill & 07 16 25.08 &&\\
~~~~Gaia DEC (J2015) \dotfill& -31 14 39.86 && \\
~~~~Gaia $\mu_\alpha$ $(\mathrm{mas}\,\mathrm{yr}^{-1})$ \dotfill& $-2.420 \pm 0.045$ && \\
~~~~Gaia $\mu_\delta$ $(\mathrm{mas}\,\mathrm{yr}^{-1})$ \dotfill& $2.440 \pm 0.047$  &&\\
~~~~Gaia DR2 Parallax $(\mathrm{mas})$\tablenotemark{b} \dotfill & $0.740 \pm 0.045$ & &\\
\sidehead{Stellar atmospheric properties  \tablenotemark{c}}
~~~~$\teffstar$ (K)\dotfill       &  \genevastarteff & \dartmouthstarteff & \\
~~~~$\feh$\dotfill                &  \genevastarfeh & \dartmouthstarfeh & $\mathcal{G}(0.03,0.10)$\\
~~~~$\vsini$ (\kms)\dotfill        &  \genevastarvsini & \dartmouthstarvsini &  $\mathcal{G}(40.58,0.34)$\\
~~~~$v_\mathrm{marcro}$ (\kms)\dotfill        &  \genevastarvmac & \dartmouthstarvmac & $\mathcal{G}(6.02,0.15)$ \\
\sidehead{Photometric properties}
~~~~Gaia $G$ (mag)\dotfill               & $12.58601\pm0.00019$ && \\
~~~~Gaia $BP$ (mag)\dotfill               & $12.689961\pm0.00093$ && \\
~~~~Gaia $RP$ (mag)\dotfill               & $12.396468\pm0.00072$ && \\
~~~~APASS $B$ (mag)\dotfill               &  $12.778 \pm 0.060$ && \\
~~~~APASS $g'$ (mag)\dotfill               &  $12.593\pm0.047$ &&\\
~~~~APASS $V$ (mag)\dotfill               &  $12.574\pm0.028$ && \\
~~~~APASS $r'$ (mag)\dotfill               & $12.642\pm0.029$ && \\
~~~~APASS $i'$ (mag)\dotfill               & $12.727 \pm 0.034$ && \\
~~~~2MASS $J$ (mag)\dotfill               & $12.139 \pm 0.023$ && \\
~~~~2MASS $H$ (mag)\dotfill               & $12.106 \pm 0.024$ && \\
~~~~2MASS $K_s$ (mag)\dotfill             & $12.040 \pm 0.026$ && \\
\sidehead{Stellar properties}
~~~~$\mstar$ ($\msun$)\dotfill      & \genevastarmass &  \dartmouthstarmass & $\mathcal{U}(0,10)$\\
~~~~$\rstar$ ($\rsun$)\dotfill      & \genevastarradius &  \dartmouthstarradius &  $\mathcal{U}(0,10)$\\
~~~~$\loggstar$ (cgs)\dotfill       & \genevastarlogg & \dartmouthstarlogg & \\
~~~~$\lstar$ ($\lsun$)\dotfill      & \genevastarlum & \dartmouthstarlum  &\\
~~~~Stellar oblateness $R_\mathrm{pole}/R_\mathrm{eq}$ & \genevastaroblate & \dartmouthstaroblate &\\
~~~~Line of sight inclination $I_*$\dotfill        & \genevairot & \dartmouthirot & $\mathcal{U}\cos I_* (0,1)$\\
~~~~$E(B-V)$ (mag)\dotfill & \genevaebmv &  \dartmouthebmv & $\mathcal{U}(0,0.1518)$ \tablenotemark{d}\\
~~~~Age (Gyr)\dotfill               & \genevastarage & \dartmouthstarage &  \\
~~~~Distance (pc) \dotfill           & $1307_{-62}^{+60}$ & $1307_{-41}^{+42}$ & \\
\enddata
\tablenotetext{a}{
  Where the quoted property is a free parameter in the global modeling, a prior and its range have been given. Steps are linear unless otherwise noted. \\
}
\tablenotetext{b}{
  Gaia DR2 parallax \citep{2018arXiv180409365G} with a systematic correction of $-82 \pm 33$ micro-arcseconds applied as per \citet{2018arXiv180503526S}, the uncertainties from Gaia and \citet{2018arXiv180503526S} have been added in quadrature. \\
}
\tablenotetext{c}{
  Derived from the global modelling described in Section~\ref{sec:analysis}, co-constrained by spectroscopic stellar parameters and the Gaia DR2 parallax.\\
}
\tablenotetext{d}{
  Uniform distribution for reddening up to the local maximum set by \citet{2011ApJ...737..103S}\\
}

\end{deluxetable*}



\begin{deluxetable*}{lrrr}
\tablewidth{0pc}
\tabletypesize{\scriptsize}
\tablecaption{
    Orbital and planetary parameters 
    \label{tab:planetparam}
}
\tablehead{ \\
    \multicolumn{1}{c}{~~~~~~~~Parameter~~~~~~~~}   &
    \multicolumn{1}{c}{\textbf{Rotational model}} &
    \multicolumn{1}{c}{Non-rotational model}  &
    \multicolumn{1}{c}{Priors}  
}
\startdata
\sidehead{\Lc{} parameters}
~~~$P$ (days)             \dotfill    & \genevaperiod & \dartmouthperiod & $\mathcal{U}(0,\infty)$\\
~~~$T_c$ (${\rm BJD}$)    
      \tablenotemark{a}   \dotfill    & \genevatzero & \dartmouthtzero & $\mathcal{U}(-\infty,\infty)$\\
~~~$T_{14}$ (days)
      \tablenotemark{a}   \dotfill    & \genevatdur & \dartmouthtdur &\\
~~~$\arstar$              \dotfill    & \genevaars & \dartmouthars & \\
~~~$\rpl/\rstar$          \dotfill    & \genevarratio & \dartmouthrratio & $\mathcal{U}(0,1)$\\
~~~$b \equiv a \cos i/\rstar$
                          \dotfill    & \genevab & \dartmouthb &\\
~~~$i$ (deg)              \dotfill    & \genevainc & \dartmouthinc & $\mathcal{U}(0,180)$\\
~~~$|\lambda|$ (deg)      \dotfill    & \genevalambda & \dartmouthlambda & $\mathcal{U}(-90,90)$\\
\sidehead{Limb-darkening and gravity darkening coefficients \tablenotemark{b}}
~~~$a_g$ (linear term)        \dotfill    & 0.3359 && \\
~~~$b_g$ (quadratic term)     \dotfill    & 0.3840 &&\\
~~~$a_r$                \dotfill    & 0.1989 &&\\
~~~$b_r$                \dotfill      & 0.3618 &&\\
~~~$a_I$                 \dotfill    & 0.1292  && \\
~~~$b_I$                 \dotfill    & 0.3321 &&\\
~~~$a_i$                 \dotfill    & 0.1507 && \\
~~~$b_i$                  \dotfill    & 0.3379 &&\\
~~~$a_{Ks}$                 \dotfill    & 0.0215&&\\
~~~$b_{Ks}$                  \dotfill    & 0.2250 && \\
~~~$\beta$ Gravity darkening exponent \dotfill & 0.1975 && \\
\sidehead{RV parameters}
~~~$K$ (\ms)              \dotfill    & \genevaKrv & \dartmouthKrv &\\
~~~$\sqrt{e} \cos\omega$ \tablenotemark{e}
                          \dotfill    & \genevartecosw & \dartmouthrtecosw & $\mathcal{U}(-1,1)$ \\
~~~$\sqrt{e} \sin\omega$ \tablenotemark{e}
                          \dotfill    & \genevartesinw & \dartmouthrtesinw & $\mathcal{U}(-1,1)$\\
~~~$e$                    \dotfill    & \genevaecc & \dartmouthecc & \\
~~~RV jitter (Coralie) (\ms)\tablenotemark{c}        
                          \dotfill    & \genevajitterA & \dartmouthjitterA & $\mathcal{U}(0,\infty)$\\
~~~RV jitter (FEROS) (\ms)\tablenotemark{c}        
                          \dotfill    & \genevajitterB & \dartmouthjitterB & $\mathcal{U}(0,\infty)$\\
~~~Systemic RV (Coralie) (\kms)\tablenotemark{d}        
                          \dotfill    & \genevagammaA & \dartmouthgammaA  & $\mathcal{U}(-\infty,\infty)$\\
~~~Systemic RV (FEROS) (\kms)\tablenotemark{d}        
                          \dotfill    & \genevagammaB & \dartmouthgammaB & $\mathcal{U}(-\infty,\infty)$\\
\sidehead{Planetary parameters}
~~~$\mpl$ ($\mjup$)\tablenotemark{f}       \dotfill    & \genevaplanetmass & \dartmouthplanetmass & $\mathcal{U}(0,\infty)$\\
~~~$\rpl$ ($\rjup$)       \dotfill    & \genevaplanetradius &  \dartmouthplanetradius  &\\
~~~$\rhopl$ (\gcmc)       \dotfill    &  \genevaplanetrho & \dartmouthplanetrho &\\
~~~$\log g_p$ (cgs)       \dotfill    & \genevaplanetgrav & \dartmouthplanetgrav &\\
~~~$a$ (AU)               \dotfill    & \genevaa & \dartmoutha &\\
~~~$T_{\rm eq}$ (K)       \dotfill    & \genevaplanetteq & \dartmouthplanetteq &\\
~~~$\Theta$\tablenotemark{g}\dotfill  & \genevaplanetSaf & \dartmouthplanetSaf&\\
~~~$\langle F \rangle$  (\ergscmsq) \tablenotemark{h} & \genevaplanetFincident & \dartmouthplanetFincident &\\
\enddata
\tablenotetext{a}{
    \ensuremath{T_c}: Reference epoch of mid transit that minimizes the
    correlation with the orbital period. BJD is calculated from UTC.
    \ensuremath{T_{14}}: total transit duration, time between first to
    last contact;
}
\tablenotetext{b}{
        Values for a quadratic law given separately for each of the filters with which photometric observations were obtained.  These values were adopted from the
        tabulations by \citet{Claret:2011} according to the
        spectroscopic an initial estimate of the stellar parameters. The limb darkening coefficients are held fixed during the global modelling.
}
\tablenotetext{c}{
    The RV jitter term is fitted for as per \citet{2016MNRAS.457.3637H} within the global modeling.
}
\tablenotetext{d}{
    The systemic RV for the system as measured relative to the telluric lines 
}
\tablenotetext{e}{
    Solutions leading to $e>1$ are rejected in the MCMC
}

\tablenotetext{f}{
    The mass measurement is quoted as the median of the posterior, with the uncertainties defined as the 68 percentile region.
}

\tablenotetext{g}{
    The Safronov number is given by $\Theta = \frac{1}{2}(V_{\rm
    esc}/V_{\rm orb})^2 = (a/\rpl)(\mpl / \mstar )$
    \citep[see][]{2007ApJ...671..861H}.
}
\tablenotetext{h}{
    Incoming flux per unit surface area, averaged over the orbit.
}
\end{deluxetable*}

\subsection{Blend scenarios}

Some stellar eclipsing binary configurations may mimic or dilute the transit and radial velocity signal of a planetary system. The Doppler tomographic detection of the planetary transit demonstrates that the orbiting companion is indeed transiting the rapidly rotating A star, and the transit signal is not originating from a background eclipsing binary blended with the bright A star \citep{2010MNRAS.407..507C}. Significant dilution of the primary transit due to a blended background star can also be ruled out. The depth of the Doppler tomographic shadow matches model predictions based on the photometric transit, and no residuals are visible after subtraction of the model transit (see residual panel of Figure~\ref{fig:dt}). We found no other sources within $5\arcsec$ in the Gaia DR2 catalog, as well as no companions within $2\arcsec$ from the Astralux lucky imaging observations. We also note that the agreement between the SED and a single star spectral model rules out any blending against a bright background star. 

Line profiles measured from least-square deconvolutions can reveal close-companions with flux ratios of $\sim 1\%$ (see Figure~5 of \citealt{2018AJ....155...35S}). To search for fainter stellar blends, we inject and attempt to recover an additional stellar signal in the out-of-transit Magellan/MIKE line profiles. We restrict the exercise to hypothetical companion stars rotating slower than the instrument resolution, such that the width of the companion line profile is fixed, and the line profile can be approximated by a Gaussian representing the instrument resolution. We simultaneously fit for the line profile of \hatcurb{} and the injected companion via an MCMC routine utilizing the \emph{emcee} package. We find that stellar companions of flux ratio $>0.008$ can be identified successfully with this exercise. No such stellar companions were found.

\section{Discussion}
\label{sec:discussion}

\subsection{System obliquity}

We were able to constrain the full 3D obliquity of the \hatcur{} system. The projected obliquity of \hatcurb{} was measured to be $|\lambda| = \genevalambda\, ^\circ$ from our spectroscopic transit observation. The line of sight inclination $I_\star$ was constrained by light curve modeling of the gravity darkening effect and constraints on physical spin-rates of the host star via isochrone modeling, and was found to be $\genevairot \, ^\circ$. Given that we know $\lambda$, $I_\star$, and the transit inclination $i$, we can calculate the true obliquity $\psi$ of the system to be $13.2_{-5.9}^{+6.4}\,^\circ$ or $160.4_{-5.2}^{+5.4}\,^\circ$ (due to the degeneracy between $\lambda$ and $\pi - \lambda$ \citealt{2009ApJ...696.1230F}).

\begin{figure}
\plotone{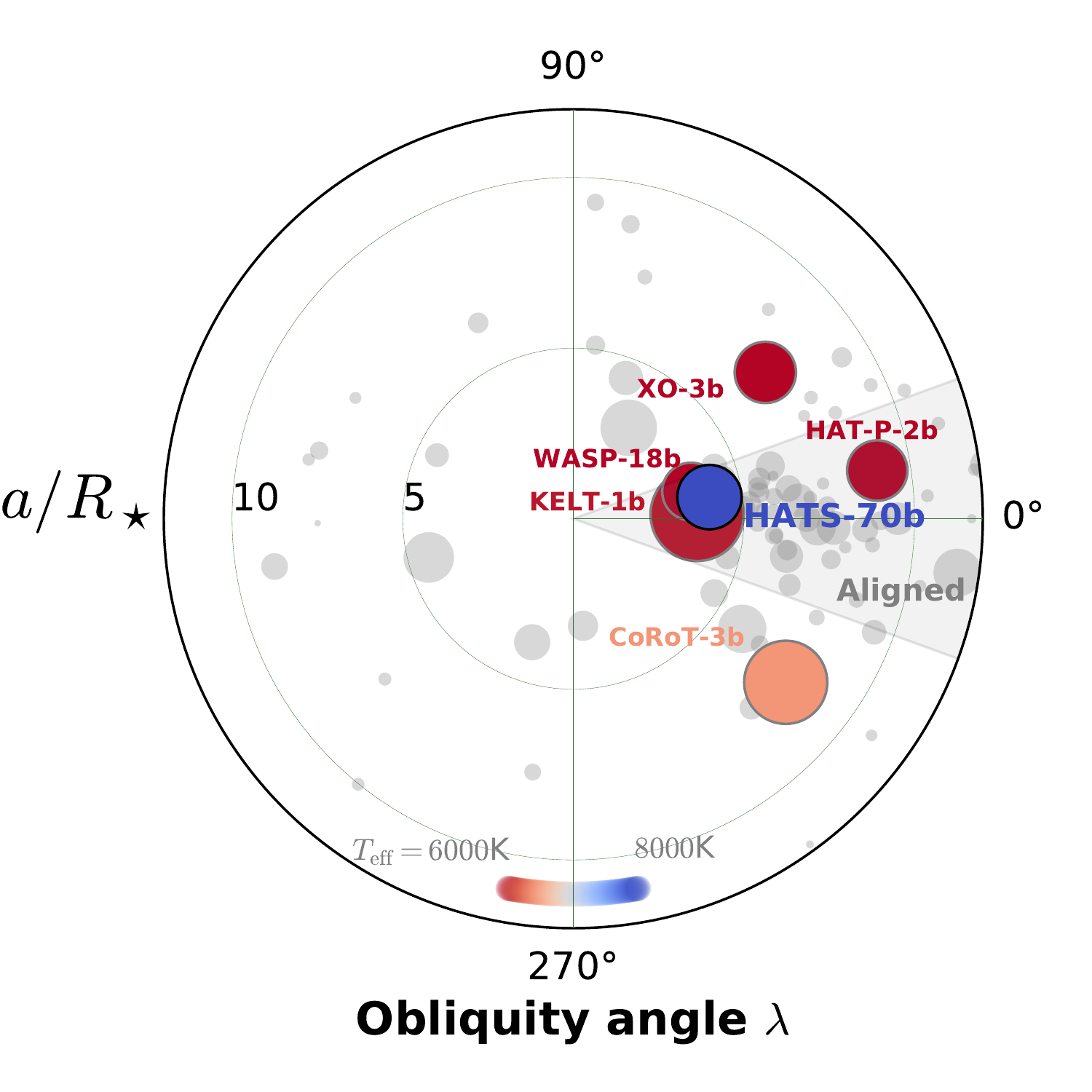}
\caption{
The obliquities of massive planets and brown dwarfs. We plot the projected obliquities of companions within the mass range of $10 < M_p < 80 \,\mjuplong$ as a function of their orbital distance from the host star $a/R_\star$. Systems orbiting F stars are colored in red, orbiting A stars in blue. All six systems for which obliquities have been measured orbit stars hotter than the Kraft break. The size of the points correspond with the mass of the transiting companion. The remainder population of transiting exoplanets for which obliquities have been measured are plotted in grey in the background. Obliquity data from TEPcat, \url{http://www.astro.keele.ac.uk/jkt/tepcat/obliquity.html}. Figure inspired by similar plots from J. Winn. 
\label{fig:lambda_ars}}
\end{figure}

\hatcurb{} is one of few massive planet / brown dwarf systems with its obliquity measured. \citet{2011A&A...533A.130H} and \citet{2017haex.bookE...2T} noted that objects more massive than $\sim 3\,\mjuplong$ tend to be in well aligned orbits. Figure~\ref{fig:lambda_ars} plots the obliquities of all systems hosting $10 < M_p < 80 \,\mjuplong$ companions, demonstrating the lack of significantly misaligned systems within this mass range. Of which, HAT-P-2b \citep{2007ApJ...665L.167W,2008A&A...481..529L,2012ApJ...757...18A}, WASP-18b \citep{2010A&A...524A..25T}, KELT-1b \citep{2012ApJ...761..123S}, and \hatcurb{} are all found to be well aligned to their host stars, while CoRoT-3b \citep{2009A&A...506..377T} and XO-3b \citep{2008A&A...488..763H,2009ApJ...700..302W,2011PASJ...63L..57H} are both orbiting in the prograde direction with the orbit normal within $40^\circ$ of the projected stellar rotation axis. In all six systems, the host stars reside above the Kraft break, where high obliquity hot Jupiters are found regularly \citep{2010ApJ...718L.145W,2012ApJ...757...18A}. For comparison, out of the lower mass $(M_p < 3\,\mjuplong)$ sample of systems definitively above the Kraft break $(T_\mathrm{eff} > 6500\,\mathrm{K})$, 5 of 11 are misaligned at angles over $40^\circ$. 

Though orbital obliquities can reflect the migrational history of systems, the spin-orbit state of the planetary system can often be modified by star-planet tidal interactions.  Tides are stronger between these massive companions and the host star than the typical hot Jupiter systems. If significant tidal evolution has occurred for the system to modify the spin orientation of the host star, then we should expect the spin rate of the star to be synchronized with the orbital period of the planet. If \hatcur{} was spin synchronized with the orbit of the brown dwarf companion, then we should observe a rotational velocity of $50.4 \pm 1.8\, \kms$, more than $5\sigma$ different from our measured rotational velocity. We can further check to see if \hatcur{} has fallen into a Darwin-stable regime by comparing the total angular momentum of the system to a critical angular momentum. Following the procedure from \citet{2010ApJ...725.1995M}, we find that the total angular momentum of the \hatcur{} system falls short of the critical value required for Darwin-stability by 0.95. In addition, the orbital angular momentum of the brown dwarf is only larger than the spin angular momentum of the system by a factor of 2.4, falling short of that required for reaching a Darwin-stable regime. For comparison, \citet{2012ApJ...761..123S} found the brown dwarf KELT-1 to be within the Darwin stable regime, and with its measured rotational velocity consistent with that expected from spin-orbit synchronization.  We can also estimate the timescale for \hatcurb{} to modify the spin of its host star using Equation 3. of \citet{2012ApJ...757....6H}, where tidal dissipation is calibrated with respect to stellar mass. In this framework, \hatcurb{} has no influence on the spin of the host star at its current orbital separation, with such modification timescales being well above the Hubble time.   Similarly,  HAT-P-2b and CoRoT-3b orbit too distant from their host stars for tidal effects to partake in the spin evolution of the host star. We can also check for signs of spin-orbit synchronization in these systems. With the exception of KELT-1, all these systems have $\vsini$ values differing from the expected synchronous rotational velocities by more than $3\sigma$, suggesting that they are not in the synchronous state.  

\citet{2014MNRAS.439.2781M} notes that systems hosting brown dwarfs $<42.5\,\mjuplong$ follow a distinct eccentricity trend, where the eccentricity envelope falls off as a function of companion mass. They note that systems with higher mass brown dwarfs do not follow this mass-dependent eccentricity envelope. As they pointed out, planet scattering simulations by \citet{2008ApJ...686..621F} predict such a mass-dependent envelope, with massive inner brown dwarfs requiring equally massive outer companions for planet-planet scattering. \citet{2014MNRAS.439.2781M} argues the necessity to form multiples of such massive objects via core accretion within a neighboring part of the protoplanetary disk biases against scattering migration of massive planets and brown dwarfs. The lack of high obliquity, high mass transiting companions tentatively agrees with this assessment. 

We note that in addition to these six systems, a number of brown dwarf-hosting stars are rapidly rotating and suitable for further spectroscopic obliquity characterizations, including CoRoT-15 \citep{2011A&A...525A..68B}, WASP-30 \citep{2011ApJ...726L..19A}, Kepler-39 \citep{2011A&A...533A..83B,2015A&A...575A..85B}, and WASP-128 \citep{2018arXiv180707557H}. We caution that of these systems, CoRoT-15 and WASP-30 are have rotational velocities similar to that expected from spin-orbit synchronization, and may have already undergone significant tidal interactions modifying the spins of the host star. Establishing a population of higher mass companions with obliquities measured, around hot stars for which planet-star tidal effects are minimal, is another pathway to determining the origins of close-in brown dwarfs.

\subsection{\hatcurb{} on the mass radius diagram}

\begin{figure*}
\plotone{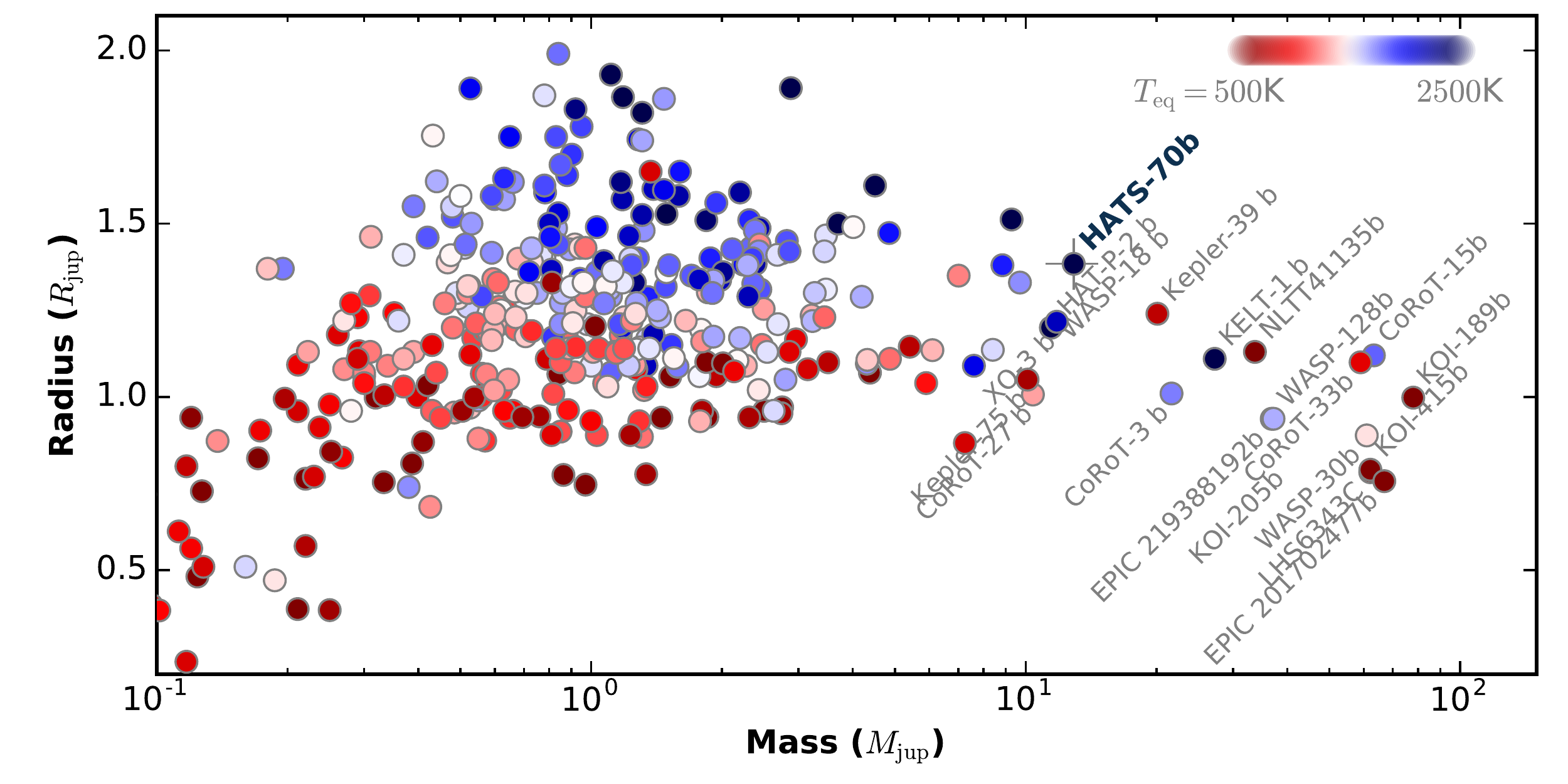}
\caption{
The mass radius diagram of transiting giant planets and brown dwarfs. The systems are colored according to their equilibrium temperatures $T_{eq}$. \hatcurb{} is amongst the most inflated of the massive planets discovered. The planet population $(M_p<30\mjuplong)$ are obtained from NASA Exoplanet Archive, while the brown dwarf population is drawn from \citet{2017AJ....153...15B}, with the addition of more recent discoveries.
Objects more massive than $10\,\mjuplong$ are labelled in grey, including
CoRoT-3b \citep{2008A&A...491..889D},
CoRoT-15b \citep{2011A&A...525A..68B},
CoRoT-27b \citep{2014A&A...562A.140P},
CoRoT-33b \citep{2015A&A...584A..13C},
EPIC 201702477b \citep{2017AJ....153...15B},
EPIC 219388192b \citep{2017AJ....153..131N},
HAT-P-2b \citep{2007ApJ...670..826B},
KELT-1b \citep{2012ApJ...761..123S},
KOI-189b \citep{2014A&A...572A.109D},
KOI-415b \citep{2013A&A...558L...6M},
Kepler-39b \citep{2011A&A...533A..83B},
Kepler-75b \citep{2013A&A...554A.114H},
LHS6343C \citep{2011ApJ...730...79J},
NLTT 41135b \citep{2010ApJ...718.1353I},
WASP-18b \citep{2009Natur.460.1098H},
WASP-30b \citep{2011ApJ...726L..19A},
WASP-128b \citep{2018arXiv180707557H},
and XO-3b \citep{2008ApJ...677..657J}.
\label{fig:massradius}}
\end{figure*}

With a radius of $R_p = \genevaplanetradius \, \rjuplong$, \hatcurb{} is amongst the largest in radius of the massive planets and brown dwarfs, and is larger than the model predictions of \citet{2007ApJ...659.1661F} by  20\,\% at 1\,Gyr. The radius excess of hot Jupiters has been well studied in the literature, with effects such as enhanced atmospheric metallicity and opacity \citep[e.g.][]{2007ApJ...661..502B,2011ApJ...736...47B}, and Ohmic dissipation \citep[e.g.][]{2010ApJ...714L.238B} being explored over the past two decades. Though no universal mechanism has emerged, the models all predict a lack of radius excess for the most massive planets and brown dwarfs. The radii of these objects are strongly defined by electron degeneracy, and are not predicted to deviate above $\approx 1\,\rjuplong$ after $\sim 500\,\mathrm{Myr}$ \citep[e.g.][]{2003A&A...402..701B,2007ApJ...659.1661F}. Figure~\ref{fig:massradius} shows the mass-radius distribution of planets and brown dwarfs against their equilibrium temperature. 

Empirically, the radius excess of hot Jupiters has long been linked to their irradiation and host star metallicity \citep[e.g.][]{2012A&A...540A..99E}, and more recent studies \citep{2016arXiv160700322B} using a much larger sample of transiting hot Jupiters found that the radii of massive planets $M_p > 2\,\mjuplong$ are also dependent on irradiation. Though an outlier, \hatcurb{} is not the only heavily irradiated high mass companion with an inflated radius. Similar to \hatcurb{}, Kepler-13b orbits an A star at a period of 1.76\,days, with a mass of $M_p = 4.94-8.09\,\mjuplong$ and a radius of $R_p = 1.406 \pm 0.038\,\rjuplong$ \citep{2011AJ....142..195S,2014ApJ...788...92S}. The radius anomaly is clearly present at least for individual systems in the high mass regime, and further tests of radius inflation models need to account for the peculiarities of these systems.

\acknowledgements  

We thank the reviewer for their thorough comments, especially in helping us better understand the dynamical history of this system. 
Development of the HATSouth
project was funded by NSF MRI grant NSF/AST-0723074, operations have
been supported by NASA grants NNX09AB29G, NNX12AH91H, and NNX17AB61G, and follow-up
observations receive partial support from grant NSF/AST-1108686.
Work by G.Z. is provided by NASA through Hubble Fellowship grant HST-HF2-51402.001-A awarded by the Space Telescope Science Institute, which is operated by the Association of Universities for Research in Astronomy, Inc., for NASA, under contract NAS 5-26555.
J.H.\ acknowledges support from NASA grant NNX14AE87G.
A.J.\ acknowledges support from FONDECYT project 1171208, BASAL CATA
PFB-06, and project IC120009 ``Millennium Institute of Astrophysics
(MAS)'' of the Millenium Science Initiative, Chilean Ministry of
Economy. N.E.\ is supported by BASAL CATA
PFB-06. R.B.\ and N.E.\ acknowledge support from project
IC120009 ``Millenium Institute of Astrophysics (MAS)'' of the
Millennium Science Initiative, Chilean Ministry of Economy.
V.S.\ acknowledges support form BASAL CATA PFB-06.  
L.M. acknowledges support from the Italian Minister of Instruction,
University and Research (MIUR) through FFABR 2017 fund. L.M.
acknowledges support from the University of Rome Tor Vergata through
``Mission: Sustainability 2016'' fund.
The modelling in this paper was performed on the Smithsonian Institution High Performance Cluster (SI/HPC). 
This work is based on observations made with ESO Telescopes at the La
Silla Observatory.
This paper also uses observations obtained with facilities of the Las
Cumbres Observatory Global Telescope.
We acknowledge the use of the AAVSO Photometric All-Sky Survey (APASS),
funded by the Robert Martin Ayers Sciences Fund, and the SIMBAD
database, operated at CDS, Strasbourg, France.
Operations at the MPG~2.2\,m Telescope are jointly performed by the
Max Planck Gesellschaft and the European Southern Observatory.  
We thank the MPG 2.2m telescope support team for their technical assistance during observations.

\facility{Magellan:Clay (MIKE), ESO:1.2m (CORALIE), MPG:2.2m (FEROS), AAT:3.9m (IRIS2), Du Pont:2.5m}

\appendix
\section{Accounting for gravity darkening in global modeling}
\label{sec:gravdark_SED}

Rotation is an important parameter governing the evolution of stars. Rotation encourages additional mixing of hydrogen in the core and envelope, extending the lifetime and luminosity of rapidly rotating stars \citep[e.g.][]{2012A&A...537A.146E}. Rotation also induces oblateness in the shape of the stars, with the poles flattened with respect to the equator. This `gravity darkening' effect causes the poles to be hotter and more luminous than the equator \citep{1924MNRAS..84..684V,2011A&A...533A..43E,Claret:2011}. 

Gravity darkening affects the resulting system parameters in two ways. The transit shape becomes a function of the transit geometry \citep{2009ApJ...705..683B}, since the brightness of the stellar surface varies as a function of latitude. The light curve would exhibit a flattened bottom in the event of a transit through the equator of the star, while a transit through the poles will appear more `V' shaped. The transit chord length also depends on the oblateness of the star. The transit duration is longer for transits along the equator of an oblate star, and shorter for transits from pole to pole. To correctly account for the transits of an oblate, rapidly rotating star, we make use of the gravity darkening transit code \emph{simuTrans} described in \citet{2018AJ....155...13H}, a numerical integrator that accounts for the gravity darkened, oblate surface of the star as per the prescription in \citet{2009ApJ...705..683B}.

\begin{figure*}
\plotone{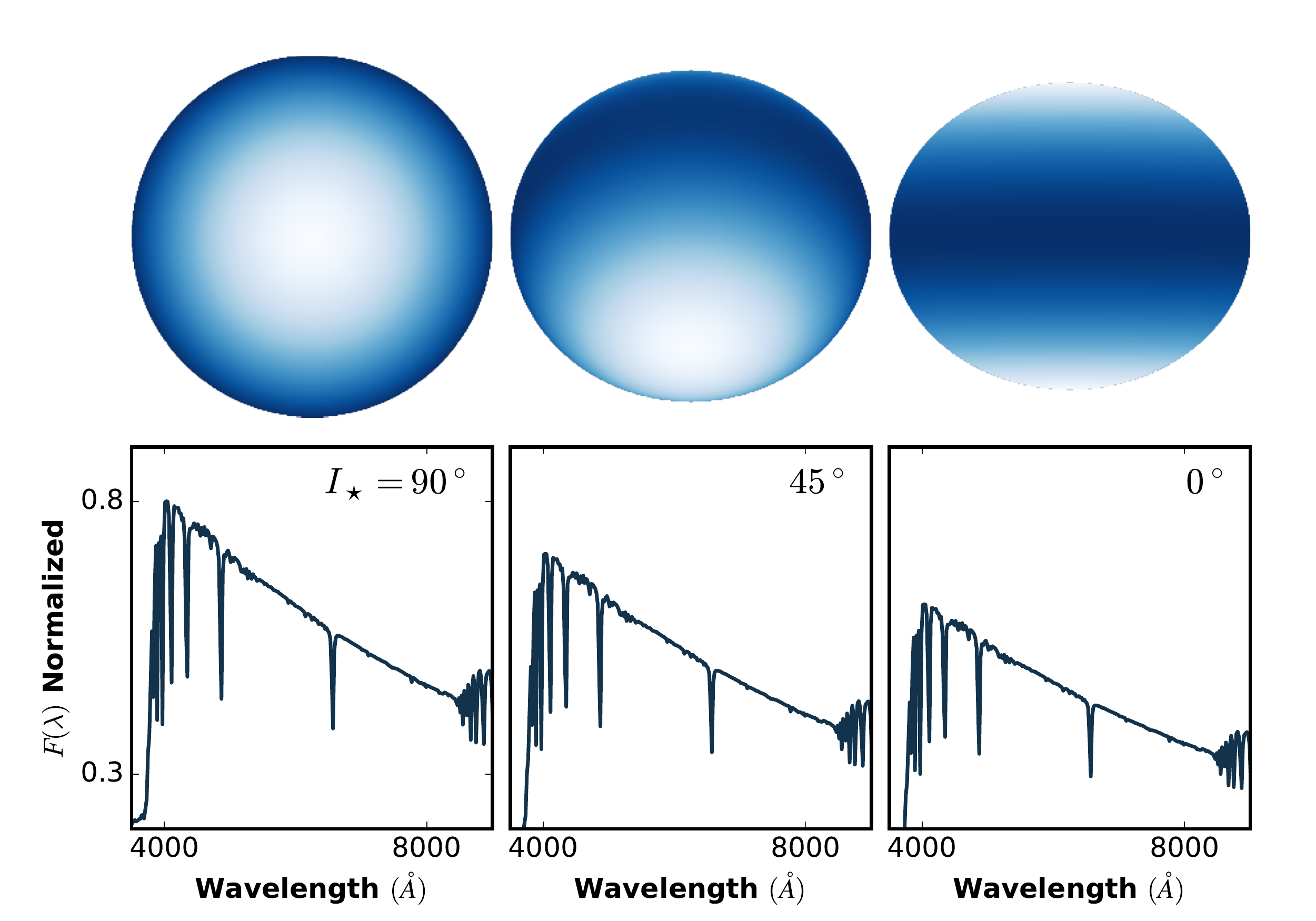}
\caption{
The apparent luminosities and temperatures of rapidly rotating stars depend on our viewing geometry. Rapidly rotating stars are gravity darkened, being brighter and hotter at the poles, and vice versa at the equator. The observed SED of the star is therefore dependent on the viewing geometry, with the star appearing bluer and brighter when viewed pole-on \textbf{(left)}, and redder and cooler when viewed equator-on \textbf{(right)}. This simulation shows a hypothetical $2.0\,\msun$, $2.0\,\rsun$ star rotating at $v_\mathrm{eq} = 200\,\kms$, with an oblateness of $R_\mathrm{pole} / R_\mathrm{eq} = 0.85$. Parameters are arbitrary and chosen to emphasize the described effect.
\label{fig:gravdark_sed}}
\end{figure*}

The effective temperature and apparent brightness of a rapidly rotating host star depends on the viewing geometry. The same star appears brighter and hotter if viewed from a polar geometry, and vice versa for the equatorial geometry (see Figure~\ref{fig:gravdark_sed}). As such, not accounting for the viewing geometry of the system can bias the derived parameters of the planet and the host star. The effect of the gravity darkening on the estimation of stellar parameters are illustrated in \citet{2015ApJ...807...58B} for the Hyades cluster. We follow their approach and adopt a modified set of Geneva 2D rotational stellar isochrones \citep{2012A&A...537A.146E}, extended to account for the influence of gravity darkening as a function of stellar inclination. For each point in the Geneva isochrones grid, we calculate a set of SEDs for stellar inclinations $(I_\star)$ between 0 and $90^\circ$ at steps of $10^\circ$. The SEDs are calculated by a disk integration of ATLAS9 fluxes \citep{Castelli:2004}, with the temperature at the stellar surface calculated using the \citet{1924MNRAS..84..684V} gravity darkening laws, assuming a gravity darkening coefficient of $\beta = 0.19$ (from \citealt{Claret:2011}, at $\teff = 8000\,\mathrm{K}$ and $\logg = 4.0$, for the Kepler band, assuming $y = 4\,\beta$). At each wavelength, the SEDs from each part of the stellar disk are summed based on a quadratic limb darkening law interpolated from \citet{Claret:2011}. We use the SEDs to calculate absolute magnitudes for the \emph{APASS} and \emph{2MASS} photometric bands for use in the eventual SED fitting. The isochrones are then interpolated via a Gradient Boosting Regression function implemented in the \emph{scikit-learn} package in \emph{Python}, along the stellar mass $(M_\star)$, radius $(R_\star)$, metallicity ([Fe/H]), equatorial rotational velocity $v$, and stellar inclination $I_\star$ plane.

\begin{figure*}
\plotone{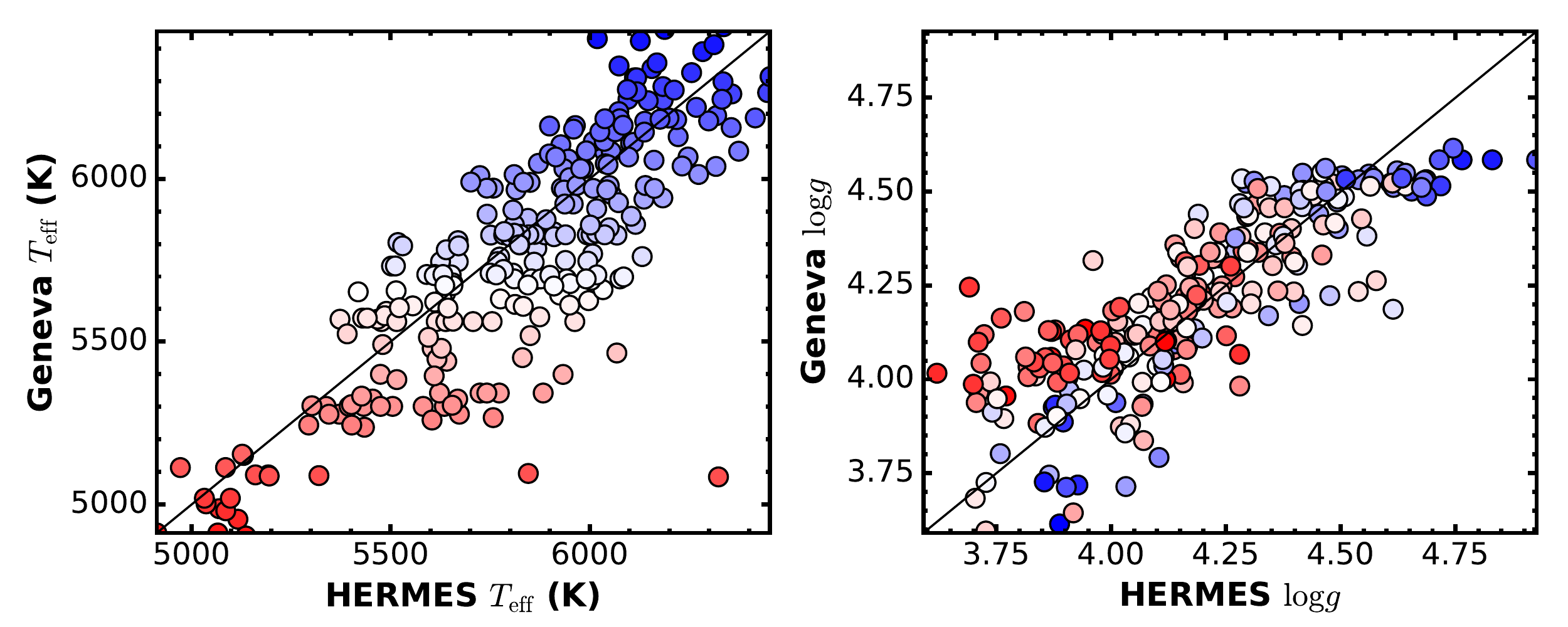}
\caption{
We test our Geneva isochrone-based SED models against a subset of randomly selected stars observed by AAT-HERMES in the TESS continuous viewing zone \citep{2018MNRAS.473.2004S}. We fit for the APASS and 2MASS magnitudes against our SED models, with their metallicity [Fe/H] fixed to that determined by AAT-HERMES. The distances to each star is constrained by their Gaia DR2 parallaxes. We find a scatter of $\sigma\,T_\mathrm{eff} = 207\,\mathrm{K}$ and $\sigma\,\log g = 0.16\,\mathrm{dex}$ in our derived effective temperature and surface gravity measurements. The points are colored by the effective temperatures of the host star, and the solid line indicates where the SED fit values agree with the AAT-HERMES values. 
\label{fig:sedtest}}
\end{figure*}

To test our interpolation, we fit the SEDs of 250 randomly selected stars sampled by the AAT-HERMES facility over the Southern ecliptic pole \citep{2018MNRAS.473.2004S}. For each star, we solve for their mass, radius, and infer their effective temperature from an SED fitting using our integrated photometric magnitudes from our SED computed via the Geneva isochrones and ATLAS9 models. The metallicities [Fe/H] are fixed to that determined spectroscopically by AAT-HERMES in this exercise. Figure~\ref{fig:sedtest} shows our SED fitting results against the spectroscopic results from \citet{2018MNRAS.473.2004S}, with a resulting scatter of $\sigma\,T_\mathrm{eff} = 207\,\mathrm{K}$ and $\sigma\,\log g = 0.16\,\mathrm{dex}$.

We note that since the metallicity is very poorly constrained in the SED fitting, in our global analysis we adopt the metallicity of the galactic disk at ages of 0.15-1.00 Gyr \citep[$\mathrm{[Fe/H]} = 0.03\pm0.12$, ][]{2003A&A...409..523R} as a Gaussian prior to help constrain the host star metallicity.

\bibliographystyle{apj}
\bibliography{hatbib}

\end{document}